\documentclass[twocolumn ,prl,longbibliography,a4paper,superscriptaddress,tightenlines,eprint,10pt]{revtex4-2}
\pdfoutput=1
\RequirePackage{silence}
\RequirePackage{float}
\setcitestyle{compressed}
\usepackage[utf8]{inputenc}
\usepackage[USenglish]{babel}
\usepackage[T1]{fontenc}
\usepackage{lmodern}
\usepackage{array,multirow}
\usepackage{algorithm}
\usepackage{algorithmic}
\usepackage{graphicx,epsf}
\usepackage[caption=false]{subfig}
\usepackage[toc,page,header]{appendix}
\usepackage{minitoc}
\usepackage{tikz}
\usetikzlibrary{quantikz}
\usepackage[bookmarksnumbered,hypertexnames=false,colorlinks,colorlinks=true,linkcolor=blue,urlcolor=black,citecolor=blue,anchorcolor=green]{hyperref}
\usepackage{bbm,braket,microtype,mathrsfs,amsmath,amssymb,color,amsthm,mathtools,fullpage,enumitem,ae,aecompl,bm,thm-restate}
\usepackage{tikz}
\usepackage{lipsum}
\usepackage{color}
\usepackage[capitalize]{cleveref}
\usepackage{silence}
\usepackage{newpxtext}
\usepackage{mathtools}

\usepackage[paper=a4paper]{geometry}

\newgeometry{top=0.75in,bottom=1.5in,left=0.75in,right=0.75in}
\usepackage[euler-digits,euler-hat-accent]{eulervm}
\allowdisplaybreaks
\newcommand{\be}{\begin{equation}}
\newcommand{\ee}{\end{equation}}
\newcommand{\ba}{\begin{eqnarray}}
\newcommand{\ea}{\end{eqnarray}}
\DeclareMathOperator{\tr}{tr}
\DeclareMathOperator{\card}{card}

\newcommand{\ignore}[1]{}
\newcommand{\st}[1]{\ket{#1}\!\!\bra{#1}}

\newcommand{\poly}[0]{\operatorname{poly}}
\newcommand{\de}[0]{{\operatorname{d}}}

\newcommand{\aver}[1]{ \left\langle  {#1}  \right\rangle }

\newcommand{\pur}{\operatorname{Pur}}

\def\norm#1{\Vert #1\Vert}
\def\CC{{\rm\kern.24em \vrule width.04em height1.46ex depth-.07ex
   \kern-.29em C}}
\def\P{{\rm I\kern-.25em P}}
\def\RR{{\rm
        \vrule width.04em height1.58ex depth-.0ex
        \kern-.04em R}}
\def\bbbone{{\mathchoice {\rm 1\mskip-4mu l} {\rm 1\mskip-4mu l}
{\rm 1\mskip-4.5mu l} {\rm 1\mskip-5mu l}}}
\def\bbbc{{\mathchoice {\setbox0=\hbox{$\displaystyle\rm C$}\hbox{\hbox
to0pt{\kern0.4\wd0\vrule height0.9\ht0\hss}\box0}}
{\setbox0=\hbox{$\textstyle\rm C$}\hbox{\hbox
to0pt{\kern0.4\wd0\vrule height0.9\ht0\hss}\box0}}
{\setbox0=\hbox{$\scriptstyle\rm C$}\hbox{\hbox
to0pt{\kern0.4\wd0\vrule height0.9\ht0\hss}\box0}}
{\setbox0=\hbox{$\scriptscriptstyle\rm C$}\hbox{\hbox
to0pt{\kern0.4\wd0\vrule height0.9\ht0\hss}\box0}}}}
\def\bbbz{{\mathchoice {\hbox{$\sf\textstyle Z\kern-0.4em Z$}}
{\hbox{$\sf\textstyle Z\kern-0.4em Z$}}
{\hbox{$\sf\scriptstyle Z\kern-0.3em Z$}}
{\hbox{$\sf\scriptscriptstyle Z\kern-0.2em Z$}}}}

\newlength{\fighskip} \fighskip=2pt
\newlength{\figvskip} \figvskip=1pt

\usepackage{hyperref}

\DeclareRobustCommand{\hypers}[2]{\texorpdfstring{\hyperlink{#1}{#2}}{#1}}
\begin{document}
\setcounter{secnumdepth}{3}
\title{Nonstabilizerness determining the hardness of direct fidelity estimation}
\author{Lorenzo Leone}\email{lorenzo.leone001@umb.edu}
\affiliation{Physics Department,  University of Massachusetts Boston,  02125, USA}
\author{Salvatore F.E. Oliviero}\email{s.oliviero001@umb.edu}
\affiliation{Physics Department,  University of Massachusetts Boston,  02125, USA}
\author{Alioscia Hamma}\email{alioscia.hamma@unina.it}
\affiliation{Dipartimento di Fisica `Ettore Pancini', Universit\`a degli Studi di Napoli Federico II, Via Cintia 80126,  Napoli, Italy}
\affiliation{INFN, Sezione di Napoli, Complesso universitario di Monte S.Angelo ed 6 via Cintia, 80126, Napoli, Italy}
%\affiliation{Physics Department,  University of Massachusetts Boston,  02125, USA}
\begin{abstract}
In this work, we show how the resource theory of nonstabilizerness quantifies the hardness of direct fidelity estimation protocols. In particular, the resources needed for a direct fidelity estimation conducted on generic states, such as Pauli fidelity estimation and shadow fidelity estimation protocols, grow exponentially with the stabilizer R\'enyi entropy. Remarkably, these protocols are shown to be  feasible only for those states that are  useless to attain any quantum speedup or advantage. This result suggests the impossibility of estimating efficiently fidelity for generic states and, at the same time, leaves the window open to those protocols specialized at directly estimating the fidelity of  particular states. We then extend our results to quantum evolutions, showing that the resources needed to certify the quality of the implementation of a given unitary $U$ are governed by the nonstabilizerness in the Choi state associated with $U$, which is shown to possess a profound connection with out-of-time order correlators.
\end{abstract}
\maketitle
\section{Introduction}
Quantum computers promise efficient solutions to problems that are otherwise intractable on classical computers~\cite{somma2008QuantumSimulationsClassical,kimble2008QuantumInternet,cirac2012GoalsOpportunitiesQuantum,bravyi2018QuantumAdvantageShallow,acin2018QuantumTechnologiesRoadmap,arute2019QuantumSupremacyUsing}. In order to fully harness the overwhelming computational advantage of quantum processors, it is first necessary to ensure their correct functioning. Unsurprisingly, the technology best suited for this task would be another quantum computer~\cite{buhrman2001QuantumFingerprinting,gottesman2001QuantumDigitalSignatures,gilyen2022ImprovedQuantumAlgorithms,cincio2018LearningQuantumAlgorithm}. Until reliable quantum technology can be realized, one must use classical resources to implement methods of {\em quantum certification}. In the last decade, there have been many attempts in tackling this problem, with a large landscape of different protocols, ranging from benchmarking~\cite{knill2008RandomizedBenchmarkingQuantum,magesan2011ScalableRobustRandomized,gambetta2012CharacterizationAddressabilitySimultaneous,magesan2012CharacterizingQuantumGates,kimmel2014RobustExtractionTomographic,wallman2014RandomizedBenchmarkingConfidence,kueng2016ComparingExperimentsFaultTolerance,roth2018RecoveringQuantumGates,wallman2018RandomizedBenchmarkingGatedependent,erhard2019CharacterizingLargescaleQuantum,onorati2019RandomizedBenchmarkingIndividual,helsen2019NewClassEfficient,helsen2019MultiqubitRandomizedBenchmarking,phillips2019BenchmarkingGaussianBoson}, quantum state~\cite{aaronson2007LearnabilityQuantumStates,cramer2010EfficientQuantumState,mari2011DirectlyEstimatingNonclassicality,gross2010QuantumStateTomography,flammia2011DirectFidelityEstimation,dasilva2011PracticalCharacterizationQuantum,dasilva2011PracticalCharacterizationQuantum,guta2018FastStateTomography,aaronson2018ShadowTomographyQuantum,takeuchi2018VerificationManyQubitStates,markham2018SimpleProtocolCertifying,pallister2018OptimalVerificationEntangled,takeuchi2019ResourceefficientVerificationQuantum,takeuchi2019ResourceefficientVerificationQuantum,elben2020CrossPlatformVerificationIntermediate,huang2020PredictingManyProperties,huang2021ProvablyEfficientMachine}, and process learning~\cite{holzapfel2015ScalableReconstructionUnitary,sekatski2018CertifyingBuildingBlocks,bouland2019ComplexityVerificationQuantum,liu2020EfficientVerificationQuantum,zhu2020EfficientVerificationQuantum,dankert2009ExactApproximateUnitary,flammia2020EfficientEstimationPauli,harper2020EfficientLearningQuantum,kliesch2019GuaranteedRecoveryQuantum,hashagen2018RealRandomizedBenchmarking} to blind computation~\cite{reichardt2013ClassicalCommandQuantum,mills2017InformationTheoreticallySecure,fitzsimons2017PrivateQuantumComputation,fitzsimons2017UnconditionallyVerifiableBlind,coladangelo2019VerifieronaLeashNewSchemes,mahadev2018ClassicalVerificationQuantuma,gheorghiu2019ComputationallySecureComposableRemote,supic2020SelftestingQuantumSystems} and quantum supremacy~\cite{arute2019QuantumSupremacyUsing,aaronson2017ComplexityTheoreticFoundationsQuantum,neill2018BlueprintDemonstratingQuantum} approaches. For a panoramic overview of the approaches within the field of quantum certification, see, e.g., Refs. \cite{kliesch2021TheoryQuantumSystem,eisert2020QuantumCertificationBenchmarking,gheorghiu2019VerificationQuantumComputation,elben2023RandomizedMeasurementToolbox,hangleiter2022ComputationalAdvantageQuantum}.
%Until the realization of reliable quantum technology, we must use classical resources to implement methods of {\em quantum certification}

A quantum certificate guarantees the correct application of a given quantum process or the correct preparation of a desired quantum state. This is commonly done in terms of a measure of quality, i.e., a measure of distance having the interpretation of worst-case distinguishability. Specifically, certifications of quantum states are phrased in terms of the fidelity between the target state $\ket{\psi}$ and the actual state $\tilde{\psi}$ prepared from the machine, while the quality of quantum gates $U$ is commonly expressed in terms of average gate fidelity~\cite{schumacher1996SendingEntanglementNoisy,nielsen2002SimpleFormulaAverage,carignan-dugas2019BoundingAverageGate,roth2018RecoveringQuantumGates}. 

The bottleneck of any quantum certification protocol is the efficiency in terms of resources. They are conventionally quantified by $(i)$ the {\em sample complexity}~\cite{kliesch2021TheoryQuantumSystem,eisert2020QuantumCertificationBenchmarking,hangleiter2022ComputationalAdvantageQuantum}, i.e., the minimal number of experiments and resulting samples that need to be prepared for a protocol to be successful, and $(ii)$ the \textit{classical postprocessing complexity}, i.e., the number of classical resources spent for postprocessing data. In particular, a protocol is said to be efficient if its total complexity scales polynomially in the number of qubits $n$; conversely, a protocol is inefficient if its complexity scales exponentially in $n$.

In this paper, we point out a very striking fact: the complexity of direct fidelity estimation protocols aimed at certifying generic quantum states is exactly quantified by the amount of {\em nonstabilizerness} in the state. Nonstabilizerness is an expensive, but fundamental fuel for quantum computation~\cite{campbell2010BoundStatesMagic,campbell2011CatalysisActivationMagic,veitch2014ResourceTheoryStabilizer,howard2017ApplicationResourceTheory,ahmadi2018QuantificationManipulationMagic,wang2019QuantifyingMagicQuantum,seddon2019QuantifyingMagicMultiqubit,liu2020EfficientVerificationQuantum,seddon2021QuantifyingQuantumSpeedups,white2021ConformalFieldTheories,qassim2021ImprovedUpperBounds,leone2022StabilizerRenyiEntropy,hahn2022QuantifyingQubitMagic,oliviero2022MeasuringMagicQuantum,haug2023ScalableMeasuresMagic}: without nonstabilizerness, a quantum computer can do nothing more than a classical computer. While simulations of stabilizer states (stabilizer resources) and Clifford circuits (stabilizer operations) are efficient on classical computers, the injection of $t$ non-Clifford gates makes the simulation exponentially harder in $t$, thus unlocking quantum advantage. Resource theory of nonstabilizerness has been widely studied and found copious applications in the broad field of fault-tolerant quantum computation~\cite{shor1996FaulttolerantQuantumComputation,gottesman1998TheoryFaulttolerantQuantum,kitaev2003FaulttolerantQuantumComputation,campbell2017RoadsFaulttolerantUniversal}, as well as classical algorithms for simulations of quantum computing architectures~\cite{gottesman1998HeisenbergRepresentationQuantuma,aaronson2004ImprovedSimulationStabilizer,bravyi2012MagicstateDistillationLow,bravyi2016ImprovedClassicalSimulation,bravyi2016TradingClassicalQuantum,bravyi2019SimulationQuantumCircuits}.

In this paper, we prove that the complexity of direct verification protocols scales exponentially with the nonstabilizerness and thus exponentially in the number of non-Clifford gates needed for the state preparation. This result implies that the certification protocol is efficient only as long as the amount of non-Clifford gates used is $\mathcal{O}(\log_2 n)$. Remarkably, this is the same threshold for a quantum state to be efficiently simulated classically~\cite{bravyi2016ImprovedClassicalSimulation}. As a consequence,  when quantum computation is able to unlock quantum speedup, then for this process direct fidelity estimation protocols are not feasible. In other words, the same complexity that makes quantum technology powerful is the one that inhibits its certification. This is the main conceptual contribution of this work.

Along these lines, we extend our results to the certification of quantum processes via direct average gate fidelity estimation. We show that the sample complexity, i.e., the number of uses of a given $U$, is quantified by multipoints out-of-time-order correlators (OTOCs) associated with the target unitary operator $U$. OTOCs are conventionally employed to probe quantum chaos: a quantum evolution is commonly considered to be chaotic in terms of attaining the Haar value for general OTOCs~\cite{bravyi2019SimulationQuantumCircuits,oliviero2021RandomMatrixTheory,leone2021IsospectralTwirlingQuantum,leone2021QuantumChaosQuantum}, that is, the value that would be reached by a random unitary operator. We claim the closer these correlators are to the Haar value, the more chaotic the evolution~\cite{bravyi2019SimulationQuantumCircuits} and the more inefficient the quantum verification. Quantum chaos is quantum---it requires an extensive quantity $\mathcal{O}(n)$ of non-Clifford resources ---and therefore it hinders quantum certification.

The paper proceeds in the following way: in Sec.~\ref{Sec: mainresult} we give an overview of the problem and informally introduce the main result of the paper. Section~\ref{Sec: tools} is devoted to the introduction of the main tools used throughout the paper. In particular, in Sec.~\ref{Sec: SRE}, we introduce the resource theory of stabilizer R\'enyi entropy, which turns out to have a deep connection to quantum fidelity estimation protocols. In Sec.~\ref{Sec: strongsimulation}, we present the algorithm for classical simulations of Clifford circuit containing a finite number of non-Clifford gates, useful in proving the main result of the paper later presented in Sec.~\ref{Sec: formalresult}. In particular, in Sec.~\ref{Sec: DFE}, we introduce the Pauli fidelity estimation protocol and bound its complexity with the stabilizer entropy, while in Sec.~\ref{Sec: SFE}, we turn to analyzing the shadow fidelity estimation protocol and show how its complexity scales exponentially with the number of non-Clifford gates, and thus with the nonstabilizerness of a given state $\ket{\psi}$. Finally, in our conclusion, we summarize the main findings of the paper and sketch ideas for future directions.

%%%%%%%%%%%%%%%%%%%%%%%%%%%%%%%%%%%%%%%%%%%%%%%%%%%%%%%%%%%%%%%%%%%%%%%%%%%%%%%%%%%%%%%%%%%%%%%%%%%%%%%%%%%%%%%%%%%%%%%%%%%%%%%%%%%%%%%%%%%%%%%%%%%%%%%%%%%%%%%%%%%%%%%%%%%%%%%%%%%%%%

\section{Fidelity estimation as a quantum certificate: statement of the main result}\label{Sec: mainresult}
Let $\psi\equiv\st{\psi}$ be the state one wants to prepare on a quantum processor and let $\tilde{\psi}$ be the state actually prepared by the quantum processor. The question behind the whole theory of quantum certification is how can one certify to what extent $\tilde{\psi}\sim \psi$ and how costly certification is? One of, if not the, most intuitive way to quantify the quality of the realization of the prepared state is to \textit{measure} the probability that $\tilde{\psi}$ is $\psi$, i.e., measure the \textit{fidelity} between $\tilde{\psi}$ and $\psi$, defined as
\be
\mathcal{F}(\ket{\psi},\tilde{\psi}):=\tr(\psi\tilde{\psi})\,.
\label{fidelity}
\ee
Operationally, the fidelity $\mathcal{F}$ quantifies the probability that $\tilde{\psi}\mapsto\st{\psi}$ and $\mathcal{F}(\ket{\psi},\tilde{\psi})=1$ if and only if $\tilde{\psi}=\st{\psi}$. In this work, we refer to \textit{direct fidelity estimation} as a protocol aimed to directly measure the fidelity $\mathcal{F}$ within an additive error $\epsilon$ [indeed an $\epsilon=\mathcal{O}(1)$ error is sufficient for a quantum certification scope since we want $\mathcal{F}\simeq 1$]. The most direct method to measure the fidelity is to measure the state $\tilde{\psi}$ in the basis in which $\ket{\psi}$ is diagonal. In other words, one can access $\mathcal{F}$ by measuring the positive operator valued measurement (POVM) given by the following set $\mathcal{S}_{\psi}=\{\st{\psi},\bbbone-\st{\psi}\}$. Unfortunately, for generic states measuring the set $\mathcal{S}_{\psi}$ is as much difficult, and noisy, as preparing the state $\ket{\psi}$. One, maybe appealing, alternative is provided by the \text{swap test}\cite{buhrman2001QuantumFingerprinting,gottesman2001QuantumDigitalSignatures}, i.e., a quantum algorithm aimed to measure the fidelity between two states, say $\ket{\psi}$ and $\tilde{\psi}$. The algorithm uses an ancillary qubit in the state $\propto\ket{0}+\ket{1}$ as the qubit control of a swap operator $T$ acting between $\ket{\psi}$ and $\tilde{\psi}$, and then measured in the basis $\ket{0}\pm \ket{1}$. The described protocol is efficient in terms of resources: a user must prepare the states $\ket{\psi}$ and $\tilde{\psi}$ an $\mathcal{O}(\epsilon^{-2})$ number of times to access the fidelity within an error $\epsilon$. As the reader might be already aware, the problem of such protocol is not the efficiency in terms of resources, but the fact that a verifier should be able to perfectly prepare the state $\ket{\psi}$ on another quantum processor. This is to say that a quantum computer is certainly able to certify the correct functioning of another noisy quantum computer. 

Unfortunately, until the advent of a completely fault-tolerant quantum technology, one must opt for other strategies. For direct fidelity estimation protocols, the rules of the game are $(i)$ the state $\ket{\psi}$ is a theoretical state, efficiently saved in a classical memory, $(ii)$ a verifier must measure the fidelity in Eq.~\eqref{fidelity} of the state $\tilde{\psi}$ by having access to $N_{\tilde{\psi}}$ state preparations of $\tilde{\psi}$ and $(iii)$ by using $N_{cl}$ resources for classical postprocessing on each $\tilde{\psi}$. We define the number of resources $\mathcal{N}$---i.e., the total \textit{complexity} of the protocol---necessary to estimate the fidelity within an error $\epsilon$ as the product of the, so-called, sample complexity $N_{\psi}$ and the classical postprocessing complexity $N_{cl}$, i.e.,
\be
\mathcal{N}=N_{\psi}\times N_{cl}
\label{complexity}
\ee
and we define a protocol to be efficient iff $\mathcal{N}=\mathcal{O}(\poly(n))$. In this work, we discuss two protocols aimed to certify the correct state preparation by directly measuring the fidelity in Eq.~\eqref{fidelity}, i.e., \textit{Pauli fidelity estimation}\cite{flammia2011DirectFidelityEstimation,dasilva2011PracticalCharacterizationQuantum} and \textit{shadow fidelity estimation}\cite{aaronson2018ShadowTomographyQuantum,huang2020PredictingManyProperties,kliesch2021TheoryQuantumSystem}. These two protocols are the only two protocols introduced in the literature, beside quantum state tomography, that have the advantage to not rely on any assumption on the state $\ket{\psi}$, and general enough to work for every state. Let us briefly and informally summarize the main steps.

\emph{Definition 1 (Pauli fidelity estimation)}
Let $\Xi_{\psi}$ be a state-dependent probability distribution on the space of a complete set of local observables. An unbiased estimator $\tilde{\mathcal{F}}$ for $\mathcal{F}$ is built in the following way: $(i)$ draw $k$ observables $O_i$ according to $\Xi_{\psi}$, $(ii)$ estimate the expectation value $\braket{O_i}_{\tilde{\psi}}$ on $\tilde{\psi}$, and $(iii)$ sum them up and define $\tilde{\mathcal{F}}:=k^{-1}\sum_i \braket{O_i}_{\tilde{\psi}}$. Note that $N_{cl}=\mathcal{O}(1)$, while the sample complexity $N_{\tilde{\psi}}\simeq k\times \max_i c_{i}$, where $c_i$ is the number of shot measurements employed to estimate $\braket{O_i}_{\psi}$.

\emph{Definition 2 (Shadow fidelity estimation)} Let $\ket{\psi}$ be a quantum state, $\tilde{\psi}$ its noisy realization on a quantum hardware, and $\{\ket{x}\}$ the computational basis. $(i)$ Draw $k$ independent Clifford circuits $C_i$ uniformly at random, $(ii)$ apply them on the prepared state, $\tilde{\psi}_i=C_{i}\tilde{\psi}C_{i}^{\dag}$, $(iii)$ measure the resulting state in the computational basis and record the result $ \bar{x}_i$, and $(iv)$ run a classical estimation algorithm to compute the outcome probability $|\!\braket{\bar{x}_i|C_i|\psi}\!|^2$ for $\bar{x}_i$ the measurement result. $(v)$ Finally define $\tilde{\mathcal{F}}:=k^{-1}\sum_{i}[(d+1) |\!\braket{\bar{x}_i|C_i|\psi}\!|^2-1]$. Note that, $N_{cl}$ counts the number of resources necessary to compute $|\!\braket{\bar{x}_i|C_i|\psi}\!|^2$ for each $i$, while the sample complexity $N_{\tilde{\psi}}\simeq k$, i.e., the number of Clifford circuits sampled.

Now we are in the position to state the main result of the paper.

\emph{Theorem 1 (Informal)} The number of resources $\mathcal{N}$ for both Pauli fidelity and shadow fidelity estimation protocols scales exponentially with the stabilizer R\'enyi entropy and thus with  the number of non-Clifford gates used to prepare the state. In particular, these protocols are feasible only for those states that can be efficiently simulable on a classical computer. 

In the next section, we introduce the main tools used to prove the above statement.

\section{Tools, definitions and techniques}\label{Sec: tools}

\subsection{Stabilizer R\'enyi entropy}\label{Sec: SRE}

The stabilizer R\'enyi entropy is a recently introduced nonstabilizerness monotone~\cite{leone2022StabilizerRenyiEntropy}, which possesses the nice property to be experimentally measurable~\cite{oliviero2022MeasuringMagicQuantum}. In this section, we briefly review some useful properties to allow easy access to the main results of the paper. We also discuss the stabilizer R\'enyi entropy associated with a unitary evolution $U$, through the Choi-Jamiolkowski isomorphism, and establish a nontrivial connection with out-of-time-order correlators, which is a result of independent interest.

Let $\rho$ be a quantum state, let $\mathbb{P}(n)$ be the Pauli group on $n$ qubits, $\mathcal{C}(n)$ the Clifford group, and $d\equiv 2^n$ the dimension of the Hilbert space. The state $\rho$ can be written in the Pauli basis as $\rho=\frac{1}{d}\sum_{P\in\mathbb{P}}\tr(P\rho)P$ and we can associate a probability distribution to the coefficients of such expansion, $\Xi_{\rho}:=\{\pur^{-1}(\rho)d^{-1}\tr^{2}(P\rho)\,|\, P\in\mathbb{P}(n)\}$, where $\pur(\rho):=\tr\rho^2$. Note that $\Xi_{\rho}(P)\ge 0$ and sum to one. The $\alpha$-stabilizer R\'enyi entropy is defined as~\cite{leone2022StabilizerRenyiEntropy}
\be
M_{\alpha}(\rho):=S_{\alpha}(\Xi_{\rho})+S_{2}(\rho)-\log_2 d, 
\ee
where $S_{\alpha}(\Xi_\rho)$ is the $\alpha$-R\'enyi entropy of the probability distribution $\Xi_{\rho}$ and $S_{2}(\rho):=-\log_2\pur(\rho)$ is the quantum 2-R\'enyi entropy of $\rho$. $M_{\alpha}(\rho)$ has the following properties: $(i)$ it follows a hierarchy $M_{\alpha}(\rho)\ge M_{\alpha^\prime}(\rho)$ for $\alpha^{\prime}<\alpha$; $(ii)$ it is faithful, i.e., $M_{\alpha}(\rho)=0$ iff $\rho=\frac{1}{d}\sum_{P\in G}\phi_{P}P$, where $G\subset \mathbb{P}(n)$ is a commuting subset of $\mathbb{P}(n)$ and $\phi_{P}=\pm1$; $(iii)$ it is invariant under Clifford rotations $C$,  $M_{\alpha}(\rho)=M_{\alpha}(C\rho C^{\dag})$; $(iv)$ it is additive: $M_{\alpha}(\rho\otimes \sigma)=M_{\alpha}(\rho)+M_{\alpha}(\sigma)$; $(v)$ it is bounded $M_{\alpha}(\ket{\psi})\le \log_2 d$. We denote the stabilizer R\'enyi entropy for a pure state $\ket{\psi}$ as $M_{\alpha}(\ket{\psi})$; for pure states only, we have that $M_{\alpha}(\ket{\psi})\le\nu(\ket{\psi})$\cite{leone2022StabilizerRenyiEntropy}, where $\nu(\ket{\psi})$ is the stabilizer nullity\cite{beverland2020LowerBoundsNonClifford} of $\ket{\psi}$, defined as $\nu(\ket{\psi})=\log_2 d-\log_2 s(\ket{\psi})$ where $s(\ket{\psi}):=|\{P\,:\, |\tr(P\st{\psi})|=1\}|$. Additionally, thanks to the bound proven in~\cite{jiang2021LowerBoundTcount}, one has $M_{\alpha}(\ket{\psi})\le t$, where $t$ is the number of $T$ gates spent in the circuit that prepares $\ket{\psi}$ from $\ket{0^n}$.

The stabilizer R\'enyi entropy is defined on states, so it is natural to compute the stabilizer R\'enyi entropy of the Choi state $\ket{U}\in\mathcal{H}^{\otimes 2}$ associated with a unitary operator $U$, as $\ket{U}:=(\bbbone\otimes U)\ket{I}$, where $\ket{I}:=\frac{1}{\sqrt{d}}\sum_{i}\ket{i}\otimes \ket{i}$. Let $\Xi_{U}$ be a probability distribution whose elements are $\Xi_{U}(P,P^{\prime}):=d^{-4}\tr^{2}(PUP^{\prime}U^{\dag})$ for $P,P^\prime \in \mathbb{P}(n)$; then the following lemma holds.

\emph{Lemma \hypertarget{choistatelemma}{1}}
The stabilizer R\'enyi entropy for $\ket{U}$ reads
\be
M_{\alpha}(\ket{U})=S_{\alpha}(\Xi_{U})-2\log_2 d
\ee
See Appendix~\ref{app:choistatelemma} for the proof. Now we are ready to state one of the main results of the paper, which builds a tight connection between the nonstabilizerness of the Choi state $\ket{U}$ and OTOCs:

\emph{Lemma \hypertarget{choiotoclemma}{2}}\label{choiotoclemma}
The $\alpha$-stabilizer R\'enyi entropy of $\ket{U}$, for $1<\alpha\in\mathbb{N}$, equals the $4\alpha$-points out-of-time order correlator
\be
M_{\alpha}(\ket{U})=\frac{1}{1-\alpha}\log_2 OTOC_{4\alpha}(U)
\ee
where $OTOC_{4\alpha}:=\frac{1}{d^{2}}\sum_{P,P^{\prime}}otoc_{4\alpha}(P,P^\prime)$, where $d \times otoc_{4\alpha}(P,P^\prime):=\tr[\langle P_{2\alpha}\prod_{i=1}^{2\alpha}P^{(U)}P^{\prime}P_{i-1}P_{i}\rangle]$ with $P_{0}\equiv \bbbone$ and $\aver{\cdot}$ is the average over $P_{1},\ldots, P_{2\alpha}$.

For a proof, see Appendix~\ref{app:choiotoclemma}. The above lemma tells the meaning of the nonstabilizerness possessed by Choi states associated with unitary evolutions: the more the nonstabilizerness $M_{\alpha}(\ket{U})$, the more chaotic is the evolution generated by $U$\cite{bravyi2019SimulationQuantumCircuits}. Lastly, we show a bound with a nonstabilizerness monotone defined by unitary operators, useful in proving the main results of the paper. Let $\nu(U)$ be the unitary stabilizer nullity defined in~\cite{jiang2021LowerBoundTcount} as $\nu(U):=2\log_2 d-\log_2 s(U)$, where $s(U):=|\{P_1,P_2\,:\, |\tr(P_1UP_2U^{\dag})|=1\}|$, i.e., $s(U)$ counts the elements of a subset of the Pauli group normalized by the adjoint action of $U$. We have the following bound.

\emph{Lemma \hypertarget{choiandstabnullity}{3}}
For any $0\le \alpha<\infty$, we have
\be
M_{\alpha}(\ket{U})\le \nu(U)
\ee

The lemma easily follows from Lemma \hyperlink{choistatelemma}{1} and the bound proven in~\cite{leone2022StabilizerRenyiEntropy}, i.e., $M_{\alpha}(\ket{\psi})\le \nu(\ket{\psi})$ for any $\alpha$. The lemma also shows that the unitary stabilizer nullity $\nu(U)$ is nothing but the stabilizer nullity of the Choi state associated with $U$, i.e., $\nu(\ket{U})=\nu(U)$. 

\subsection{Strong classical simulation of states beyond stabilizer states}\label{Sec: strongsimulation}
In this section, we present a brief and simplified review of the classical simulation method for states beyond stabilizer states, that will be useful in proving Theorem~\hyperlink{thshadow}{3}.

Imagine we are given the quantum circuit $U_t$, as a Clifford circuit plus a number $t$ of $T$ gates circuit, that builds a state $\ket{\psi}\equiv U_t\ket{0^n}$ starting from a reference state $\ket{0^n}$. Throughout the paper, we refer to ``strong simulation'' as the ability to (exactly) compute the outcome probability $|\braket{x|\psi}|^2$ for some $n$-bit string $\ket{x}$. The following simulation algorithm is not a \textit{state of the art} kind of algorithm. We describe it to illustrate why and how the strong simulation of Clifford+$T$ circuits scales exponentially in $t$, keeping the technicalities as simple as possible. See e.g., Refs.~\cite{bravyi2016ImprovedClassicalSimulation,bravyi2019SimulationQuantumCircuits,kocia2022MoreOptimalSimulation,kissinger2022ClassicalSimulationQuantum,kissinger2022SimulatingQuantumCircuits} for state-of-the-art simulation algorithms. We anticipate and remark that any simulation algorithm aimed to simulate Clifford + $T$ circuits scales exponentially in the number of $T$ gates.

First of all, thanks to the Gottesman-Knill theorem, one can compute the overlap between any two $n$-qubit stabilizer states $\ket{\omega_1},\ket{\omega_2}$ as $\braket{\omega_1|\omega_2}=b2^{-p/2}e^{i\pi m/4}$, for some $b=\{0,1\}$, integer $p\in[1,n]$, and $m\in \mathbb{Z}_8$ with an algorithm having runtime $\mathcal{O}(n^3)$~\cite{aaronson2004ImprovedSimulationStabilizer}. Conversely, if one is provided with the decomposition of $\ket{\psi}$ into elementary gates of Clifford+T circuits, the simulation cost scales exponentially in the number of $T$ gates, as shown below.

The algorithm starts from the following simple observation. Define the $T$ gadget as the following state $\ket{T}\propto \ket{0}+e^{i\pi/4}\ket{1}$. The $T$ gadget, together with a controlled-NOT---here denoted as $CX_{i,j}$, where $i$ is the control and $j$ the target---and measurement in the $Z$ basis, can be spent to apply a $T$ gate. The protocol is the following. Let $\ket{\psi}$ be a $n$ qubit state on which one wants to apply a $T$ gate on the $i$th qubit for $i\in\{1,n\}$. Let $n+1$ be the labeling of the ancillary qubit corresponding to $\ket{T}$. The first thing to do is to append the $T$ gadget as $\ket{\psi}\mapsto\ket{\psi}\ket{T}$; then apply a $CX$, having control on $i$ and target on $n+1$ as $\ket{\psi}\ket{T}\mapsto CX_{i,n+1}(\ket{\psi}\ket{T})$; and then measure the $n+1$ qubit in the $\{\ket{0},\ket{1}\}$ basis. If the measurement leads to the outcome $``0''$, then $\st{0}C_{i,n+1}(\ket{\psi}\ket{T})\propto T_i\ket{\psi}\ket{0}$, while, if it leads to the outcome $``1''$, then $\st{1}C_{i,n+1}(\ket{\psi}\ket{T})\propto S_{i}^{\dag}T_i\ket{\psi}\ket{1}$. Thus \textit{adapting} the application of an $S$ gate, i.e., $S\equiv\operatorname{diag}(1,i)$,  on the $i$th qubit conditioned to the measurement result, leads to the application of a $T$ gate on the $i$th qubit. For a generic $n$-qubit $t$-doped Clifford circuit $U_t:=C_{t}T_{i_t}C_{t-1}T_{i_{t-1}}\cdots C_{1}T_{i_1}C_0$, i.e., Clifford circuits interleaved with the application of $t$ non-Clifford gates, we can define the following $(n+t)$-qubit Clifford circuit
\be
C_{U_t}=C_{t}CX_{i_t,n+1}C_{t-1}CX_{i_{t-1}, n+2}\cdots C_{1}CX_{i_1,n+t}C_0\,,
\label{gadget}
\ee
i.e., we replace all the $T_{i_k}$ gates with $CX$ gates $CX_{i_{k},n+k}$ controlling on the $i_{k}$th qubit and acting on the $k$th auxiliary qubit for $k\in\{n+1,n+t\}$. $C_{U_t}$ is called the \textit{gadgetized} version of $U_t$. Thanks to the observation described above, one can write the outcome probability as
\be
|\braket{x|U_t|0^n}|^2=2^t|\braket{x,0^{t}|C_{U_t}|0^n,T^{\otimes t}}|^2\,,
\ee
i.e., the probability that the $n$-qubit circuit $U_{t}$ acting on $\ket{0^n}$ leads to $\ket{x}$ is proportional to the probability that the \textit{gadgetized} version $C_{U_t}$ [Eq.~\eqref{gadget}] acting on $\ket{0^n}\otimes \ket{T}^{\otimes t}$  leads to $\ket{x}\otimes\ket{0^t}$. The proportionality factor $2^t$ is due to postselection~\cite{bravyi2016ImprovedClassicalSimulation}. Next, observe that one can write $\ket{T}^{\otimes t}=\sum_{y\in\{0,1\}^t}e^{i\frac{\pi}{4}hw(y)}\ket{y}$, where $hw(y)$ is the Hamming weight of the $t$-bit string $y\in\{0,1\}^t$. In other words, one can write $t$ copies of the $T$ gadget as a combination of $2^t$ computational basis states. We can thus estimate the outcome probability $|\braket{x|U_t|0^n}|^2$ as
\be
|\braket{x|U_t|0^n}|^2=2^t\left|\sum_{y\in\{0,1\}^t}e^{i\frac{\pi}{4}hw(y)}\braket{x,0^{t}|C_{U_t}|0^n,y}\right|^2\,,
\label{evaluatingsimulating}
\ee
i.e., by evaluating the overlaps between $C_{U_t}\ket{0^n,y}$ and $\ket{x,0^t}$ via the Gottesman-Knill theorem for every $y\in\{0,1\}^t$ and then summing them up.
The above method, for the exact computation of the outcome probability, leads to a classical simulation cost $\mathcal{O}(2^t(n+t)^{3})$, where the exponential scaling in $t$ comes from the fact that there are exponentially many $t$-bit strings $y$ in the sum of Eq.~\eqref{evaluatingsimulating}, while the factor $(n+t)^3$ comes directly from the Gottesman-Knill theorem.

Now we are ready to discuss the main contributions of the paper, showing that the efficiency of direct fidelity estimation protocols is governed by nonstabilizerness.

\section{Stabilizer R\'enyi entropy and fidelity estimation: formal results}\label{Sec: formalresult} 
The following section is devoted to the presentation of the main theorems of the paper in a formal fashion. Specifically, in Sec.~\ref{Sec: DFE}, we first describe Pauli fidelity estimation protocol, first introduced by~ \cite{flammia2011DirectFidelityEstimation, dasilva2011PracticalCharacterizationQuantum}, and bound the number of resources $\mathcal{N}$ with the stabilizer entropy. In Sec.~\ref{Sec: SFE}, we describe the shadow estimation protocol introduced in~\cite{aaronson2018ShadowTomographyQuantum,huang2020PredictingManyProperties} and show that the total complexity scale exponentially with the number of non-Clifford gates spent to prepare the state. Finally, in Sec.~\ref{comparisonSFEDFE}, we show that Pauli fidelity estimation performs better than shadow fidelity estimation in terms of resources. The analyzed protocols have the advantage of being problem-agnostic protocols, i.e., they do not rely on any additional assumption and work for generic states. We will demonstrate that, while these protocols have wide applicability, they are only feasible and scalable for the class of states that do not provide a quantum computational advantage.

\subsection{Pauli fidelity estimation}\label{Sec: DFE}
Here we show that the stabilizer R\'enyi entropy directly quantifies the resources needed to estimate the fidelity, the distance in $2$-norm---for pure states and mixed states---up to an accuracy $\epsilon$ and success probability lower bounded by $1-\delta$. In particular, we prove that the stabilizer R\'enyi entropy quantifies the number of resources required for a direct fidelity estimation, conducted via Monte Carlo sampling: Pauli fidelity estimation. This protocol was first introduced in \cite{flammia2011DirectFidelityEstimation, dasilva2011PracticalCharacterizationQuantum} to directly access the fidelity of a state preparation $\tilde{\psi}$ and then experimentally employed in\cite{lanyon2017EfficientTomographyQuantum,fedorov2012ImplementationToffoliGate}. 

The protocol proceeds as follows. Let $\ket{\psi}$ be the pure state one aims to prepare on a quantum processor and let $\tilde{\psi}$ be the state (in general mixed) actually prepared by the quantum processor. As discussed in Sec.~\ref{Sec: mainresult}, a measure of quality of $\tilde{\psi}$ is provided by the fidelity $\mathcal{F}$ between the theoretical state $\ket{\psi}$ and $\tilde{\psi}$, i.e.,
\be
\mathcal{F}(\ket{\psi},\tilde{\psi}):=\tr(\psi\tilde{\psi})
\label{fidelity1}
\ee
where $\psi:=\st{\psi}$. One can rewrite Eq. \eqref{fidelity1} in the Pauli basis $\mathcal{P}(n)$ as $\mathcal{F}(\ket{\psi},\tilde{\psi})=\frac{1}{d}\sum_{P}\tr(P\psi)\tr(P\tilde{\psi})$ and define $X_{P}:=\frac{\tr(P\tilde{\psi})}{\tr(P\psi)}$; note $\Xi_{\psi}:=\{\Xi_{\psi}(P)\equiv d^{-1}\tr^{2}(P\psi)\,|\,P\in\mathbb{P}(n)\}$ is the probability distribution introduced in Sec.~\ref{Sec: SRE} for $\psi$ being a pure state. Thus we can write the fidelity as an expectation value over $\Xi_{\psi}$:
\be
\mathcal{F}(\ket{\psi},\tilde{\psi})=\sum_{P}X_{P}\Xi_{\psi}(P)\equiv \aver{X_P}_{\Xi_{\psi}}
\label{average1}
\ee
i.e., the fidelity between the theoretical pure state $\psi$ and the prepared state $\tilde{\psi}$ can be recast as an average of measurable numbers $X_{P}$ on the probability distribution $\Xi_{\psi}$. Following\cite{flammia2011DirectFidelityEstimation,kliesch2021TheoryQuantumSystem}, we use the following protocol to estimate the average in Eq. \eqref{average1}: $(i)$ extract $k$ Pauli operators $P_{1},\ldots,P_{k}\in\mathbb{K}$ according to the state-dependent probability distribution $\Xi_{\psi}$; $(ii)$ for each extraction $P\in\mathbb{K}$ of the Pauli observable $P$ construct $c_{P}(\tilde{\psi})$ copies of the state $\tilde{\psi}$ to estimate the expectation value $\tr(P\tilde{\psi})$; $(iii)$ compute the unbiased estimator of the fidelity $\mathcal{F}(\ket{\psi},\tilde{\psi})$ given by $\tilde{\mathcal{F}}=\frac{1}{k}\sum_{P\in\mathbb{K}}\tilde{X}_{P}$ where
$\tilde{X}_{P}=\tr^{-1}(P\psi)c_{P}(\tilde{\psi})^{-1}\sum_{j=1}^{c_{P}(\tilde{\psi})}\mathcal{P}_{Pj}(\tilde{\psi})$ and $\mathcal{P}_{Pj}(\tilde{\psi})$ is the outcome of a one-shot measurement of the observable $P$ on the $j$th copy of $\tilde{\psi}$. We quantify the resources needed for the estimation---up to an accuracy $\epsilon$ and failure probability $\le\delta$---as the number of copies of $\tilde{\psi}$ to be prepared on the machine:
\be
N_{\tilde{\psi}}:=\sum_{P\in \mathbb{K}}c_{P}(\tilde{\psi})
\ee
Surprisingly, the total resources $N_{\tilde{\psi}}$ are exactly quantified by the nonstabilizerness of $\ket{\psi}$, measured via the stabilizer R\'enyi entropy $M_{\alpha}(\ket{\psi})$ as the next theorem states.

\emph{Theorem \hypertarget{theorempurestates}{2}}
The sample complexity $N_{\tilde{\psi}}$ needed to measure the fidelity $\mathcal{F}$ with accuracy $\epsilon$ and success probability $1-\delta$ is bounded:
\be
\frac{2}{\epsilon^{2}}\ln(2/\delta)\exp[M_{2}(\ket{\psi})]\!\!\le N_{\tilde{\psi}}\!\!\le\!\! \frac{64}{\epsilon^{4}}\ln(2/\delta)\exp[M_{0}(\ket{\psi})],
\ee
where $M_{2}(\ket{\psi})$ and $M_{0}(\ket{\psi})$ are the $2$ and the $0$-stabilizer R\'enyi entropy respectively. Then, since $N_{cl}=\Theta(1)$, one has that the number of resources obeys $\mathcal{N}_{P}=\Theta(N_{\tilde{\psi}})$.

See Appendix~\ref{app:theorempurestates} for a proof. The above theorem tells us that the more the nonstabilizerness of the quantum state one aims to prepare on the quantum machine, the harder is the verification through Pauli fidelity estimation protocol. Let us use the theorem to determine the scaling of $N_{\tilde{\psi}}$ for an important class of states, i.e., the $t$-doped stabilizer states. A $t$-doped stabilizer state, denoted as $\ket{\psi_t}$, is the output state of a circuit composed by Clifford gates doped with a finite amount $t$ of non-Clifford resources. The best classical algorithm able to simulate such states scales as $\mathcal{O}(poly(n)\exp[t])$\cite{bravyi2016ImprovedClassicalSimulation}, providing an insightful threshold for the onset of quantum advantage: as long as $t=\mathcal{O}(\log_2 n)$, such states can be efficiently simulated classically and therefore cannot provide any quantum speedup. We have the following result:

\emph{Corollary \hypertarget{averagestabresources}{1}}
The (average) number of resources to verify a $t$-doped stabilizer state $\psi_{t}$ grows exponentially in $t$:
\be
\Theta(\exp[t\log_2 4/3])\le \aver{N_{\psi_{t}}}\le\Theta(\exp[t])\le \Theta(d)
\ee
see Appendix~\ref{app:averagestabresources} for a proof. Two comments are in order here: first, the hardness of the verification of $t$-doped stabilizer states quickly saturates the bound, growing exponentially in $t$. Second, this is telling us that the above protocol is efficient only for those states with $t$ at most $\mathcal{O}(\log_2 n)$, which is useless for quantum computation. 
%%%%%%%%%%%%%%%%%%%%%%%%%%%%%%%%%%%%%%%%%%%%%%%%%%%%%%%%%%%%%%%%%%%%%%%%%%%%%%%%%%%%%%%%%%%%%%%%%%%%%%%%%%%%%%%%%%%%%%%%%%%%%%%%%%%%%%%%%%%%%%%%%%%%%%%%%%%%%%%%%%%%%%%%%%%%%%%%%%%%%%%%%%%%%%%%%%%%%%%%%%%%%%%%

Let us now extend the above results to mixed states. Suppose one aims to prepare a mixed state $\rho$ on a quantum processor. Let $\tilde{\rho}$ be the actual state prepared from the quantum machine. One way to check whether the preparation is faithful is to evaluate the difference in $2$-norm between $\rho$ and $\tilde{\rho}$\cite{odonnell2016EfficientQuantumTomography}:
\be
\norm{\rho-\tilde{\rho}}_{2}=\sqrt{\pur(\rho)}\sqrt{1+\frac{\pur(\tilde{\rho})}{\pur(\rho)}-2\Phi(\rho,\tilde{\rho})}
\label{norm2mixed}
\ee
where we defined $\Phi(\rho,\tilde{\rho}):=\frac{\tr(\rho\tilde{\rho})}{\pur(\rho)}$ as the overlap between $\rho$ and $\tilde{\rho}$. In order to evaluate the above, one needs to measure both $\Phi(\rho,\tilde{\rho})$ and $\pur(\tilde{\rho})$. Nonetheless, here we are only concerned with the overlap $\Phi(\rho,\tilde{\rho})$, because it is the only quantity involving a direct comparison between the theoretical state $\rho$ and the actual state $\tilde{\rho}$. Note that the purity $\pur(\tilde\rho)$ can be estimated efficiently by employing the standard technique of the {\em swap test}\cite{buhrman2001QuantumFingerprinting,gottesman2001QuantumDigitalSignatures}. Writing $\Phi(\rho,\tilde{\rho})$ in the Pauli basis, one can recast it in terms of the expectation value, $\Phi(\rho,\tilde{\rho})=\aver{X_{P}(\rho)}_{\Xi_{\rho}}$, of $X_{P}(\rho):=\frac{\tr(P\tilde{\rho})}{\tr(P\rho)}$, on the probability distribution $\Xi_{\rho}$ associated to the mixed state $\rho$, $\Xi_{\rho}=\{d^{-1}\pur^{-1}(\rho)\tr^{2}(P\rho)\,|\,P\in\mathbb{P}(n)\}$. Thus, following the protocol described above, we can estimate the above average by an importance sampling of the probability distribution $\Xi_{\rho}$ and construct an unbiased estimator $\tilde{\Phi}(\rho,\tilde{\rho})=\frac{1}{k}\sum_{P\in\mathbb{K}}\frac{1}{\tr(P\rho)}\frac{1}{c_{P}(\rho)}\sum_{j=1}^{c_{P}(\rho)}\mathcal{P}_{Pj}(\tilde{\rho})$,
where $c_{P}(\rho)$ are the number of copies of $\tilde{\rho}$ needed to estimate $\tr(P\tilde{\rho})$ and $\mathcal{P}_{Pj}(\tilde{\rho})$ is the outcome of the measurement of $P$ on the $j$th copy of $\tilde{\rho}$. The number of resources needed to access the overlap $\Phi(\rho,\tilde{\rho})$ is given again by the total number of copies of $\tilde{\rho}$, i.e., $N_{\tilde\rho}=\frac{1}{k}\sum_{P\in\mathbb{K}}c_{P}(\tilde\rho)$. We are now ready to bound $N_{\tilde{\rho}}$ in terms of the stabilizer R\'enyi entropy for mixed states.

\emph{Corollary 2}
The number of resources $N_{\tilde{\rho}}$ needed to measure the overlap $\Phi(\rho,\tilde{\rho})$ with an accuracy $\epsilon$ and success probability $\ge 1-\delta$ is bounded by
\be\label{eq:pausam}
\frac{2}{\epsilon^2}\ln(2/\delta)\exp[M_{2}(\rho)]\le N_{\rho}\le\frac{64}{\epsilon^4}\ln(2/\delta)\exp[M_{0}(\rho)]
\ee
For mixed states also, the number of resources needed to measure the overlap between $\rho$ and $\tilde{\rho}$ is exactly quantified by the stabilizer R\'enyi entropy $M_{\alpha}(\rho)$. 

We remark once again that the Pauli fidelity estimation protocol described above is state-agnostic, i.e., it does not make any assumption on the nature of the state $\ket{\psi}$ and, consequently, it works for every state. Nevertheless, there is a rich literature of examples in which one can efficiently certify the preparation of a particular set of states. See for example hypergraph states~\cite{morimae2017VerificationHypergraphStates,markham2018SimpleProtocolCertifying,zhu2019EfficientVerificationHypergraph,dangniam2020OptimalVerificationStabilizer,zhou2022EntanglementRandomHypergraph}, i.e., states built from $(\ket{0}+\ket{1})^{\otimes n}$ with the application of $CCZ$ gates, and bipartite pure states~\cite{pallister2018OptimalVerificationEntangled,li2019EfficientVerificationBipartite,zhu2019EfficientVerificationPure,takeuchi2018VerificationManyQubitStates,liu2019EfficientVerificationDicke,li2021VerificationPhasedDicke}.

In what follows, we describe an extension of the Pauli fidelity estimation protocol for state-aware verifiers. Suppose a one wants to prepare the state $\ket{\psi}$ and, beside knowing the quantum circuit able to prepare $\ket{\psi}$ from a reference state $\ket{0^n}$, one knows a complete set of stabilizer observables $O_{1},\ldots, O_{d}$, i.e., Hermitian operators such that $O_{i}\ket{\psi}=\pm \ket{\psi}$ and $[O_i,O_j]=0\, \forall i,j=1,\ldots, d$. Note that every state $\ket{\psi}$ possesses one. Let $\mathbb{O}$ be a complete basis of operators such that $O_{1},\ldots, O_{d}\in\mathbb{O}$ and define the state-dependent probability distribution $\Xi_{\psi}^{\mathbb{O}}$, in the same fashion of $\Xi_{\psi}$ defined in Sec.~\ref{Sec: SRE}, as
\be
\Xi_{\psi}^{\mathbb{O}}:=\{d^{-1}\tr^{2}(O_{i}\psi), O_{i}\in \mathbb{O}\}\,.
\ee
Following the protocol described at the beginning of the section and replacing $\Xi_{\psi}\mapsto \Xi_{\psi}^{\mathbb{O}}$ and $P_i\mapsto O_i$, one can estimate the fidelity $\mathcal{F}(\ket{\psi},\tilde{\psi})$ as $\mathcal{F}(\ket{\psi},\tilde{\psi})=\sum_{O_i}\frac{\tr(\tilde{\psi} O_i)}{\tr(\psi O_i)}\Xi_{\psi}^{\mathbb{O}}(O_i)$. At this point,  one can define a entropy $S(\Xi_{\psi}^{\mathbb{O}})$ for the probability distribution $\Xi_{\psi}^{\mathbb{O}}$, similarly to the definition of stabilizer entropy in Sec.~\ref{Sec: SRE}. Since $O_{1},\ldots, O_{d}\in\mathbb{O}$, it is straightforward to verify that $S(\Xi_{\psi}^{\mathbb{O}})=n$. At this point, the following corollary readily descends from Theorem~\hyperlink{theorempurestates}{2}.
\emph{Corollary \hypertarget{cor: gDFe}{3}}
Given the knowledge of a complete set of stabilizers $O_{1},\ldots, O_d$ for a state $\ket{\psi}$ and the ability to perform measurements in the basis of operators $\mathbb{O}\ni O_{1},\ldots, O_d$, the number of resources $\mathcal{N}_P$ to estimate the fidelity, via the generalized Pauli fidelity estimation protocol, between the state $\ket{\psi}$ and its noisy realization $\tilde{\psi}$ within an error $\epsilon$ and with failure probability $\delta$ is given by
\be
\mathcal{N}_{P}^{\mathbb{O}}=\Theta(\epsilon^{-2}\ln 2/\delta)
\ee

Before moving on to the next section, a couple of remarks are in order. The above corollary tells us that it is sufficient to know a set of stabilizer operators to certify any state $\ket{\psi}$ in the Hilbert space. However, the complete knowledge of a complete set of stabilizer observables is an assumption way stronger than one may think and, in practice, can be fulfilled for a restricted set of simple states. In general, a complete set of stabilizer observables requires exhaustive search in an exponentially large space to be found, as such operators are highly nonlocal. Moreover, even if, for some reason, a verifier knows a complete set of stabilizer observables, measurements in such a basis require, in general, an exponential space in classical memory to be performed. At least, the above corollary gives a simple recipe for direct fidelity estimation protocol for those states whose stabilizers can be easily found. Let us make a clarifying example: consider a single qubit state $\ket{\phi}$ and the $n$-fold tensor product $\ket{\phi}^{\otimes n}$. Let $o_1,o_2$ be two single qubit Hermitian operators such that $o_1\ket{\phi}=o_2\ket{\phi}=\ket{\phi}$. The set $\{o_1,o_2\}^{n}$ $(i)$ is a complete set of stabilizer observables for $\ket{\phi}^{\otimes n}$; $(ii)$ can be efficiently found by exhaustive search in the space of one qubit, and $(iii)$ measurements in such basis can be easily performed being a tensor product basis of operators. The same conclusions can be reached for hypergraph states, whose stabilizers are well known and easily implemented being Hermitian Clifford operators (see \cite{morimae2017VerificationHypergraphStates,zhu2019EfficientVerificationHypergraph}). In particular, for hypergraph states, Corollary~\hyperlink{cor: gDFe}{3} constitutes a simple and alternative proof of their efficient certification. 

To conclude, one could argue about the validity of the main statement of the paper: is nonstabilizerness really playing a role in the efficiency of direct fidelity estimation? The answer is ``yes'' because, in general, for state-agnostic verifiers, the best thing that one can do is to perform measurements in the Pauli basis, i.e., the native logic basis of operators, which turns out to be much more feasible than an exhaustive search in an exponentially large space.

%%%%%%%%%%%%%%%%%%%%%%%%%%%%%%%%%%%%%%%%%%%%%%%%%%%%%%%%%%%%%%%%%%%%%%%%%%%%%%%%%%%%%%%%%%%%%%%%%%%%%%%%%%%%%%%%%%%%%%%%%%%%%%%%%%%%%%%%%%%%%%%%%%%%%%%%%%%%%%%%%%%%%%%%%%%%%%%%%%%%%%%%%%%%%%%%%%%%%%%%%%%%%%%%

\subsection{Shadow Fidelity estimation}\label{Sec: SFE}
In this section, we prove that the resources needed to estimate the fidelity via the shadow estimation protocol scale with the number of non-Clifford gates used for the state preparation. Let $\ket{\psi}$ be the state to be prepared on a quantum computer and $\tilde{\psi}$ its noisy realization prepared by the hardware. The protocol proceeds in the following steps. $(i)$ Draw $C_1,\ldots, C_{k}\in \mathcal{C}(n)$ independent Clifford unitary operators. $(ii)$ For each $C_i\in\mathcal{C}(n)$, apply $\tilde{\psi}_i\equiv C_{i}^{\dag}\tilde{\psi}C_i$, measure $\tilde{\psi}_i$ in the computational basis $\{\ket{x}~|~ x\in\{0,1\}^n\}$, and record the outcome $\bar{x}_i$. $(iii)$ Perform a classical estimation of the outcome probability $|\!\braket{\bar{x}_i|C_i|\psi}\!|^2$. An unbiased estimator for the fidelity is then given by
\be
\tilde{\mathcal{F}}=\frac{1}{k}\sum_{i}[(d+1)|\!\braket{\bar{x}_i|C_i|\psi}\!|^2-1]\,,
\label{unbiasedshadow}
\ee
i.e., if $p_i[\bar{x}]\equiv\tr[\tilde{\psi}_i\st{\bar{x}}]$ is the probability that a measurement of $\tilde{\psi}_i$ gives the string $\bar{x}$, then  $\sum_{\bar{x}\in\{0,1\}^n }\braket{p_i[\bar{x}]\tilde{\mathcal{F}}}_{C_i\in\mathbb{C}}=\tr(\psi\tilde{\psi})$. A complete and detailed derivation of Eq.~\eqref{unbiasedshadow} is to be found in~\cite{huang2020PredictingManyProperties,kliesch2021TheoryQuantumSystem}. Before stating the result of this section, let us focus on step $(iii)$ of the protocol. Shadow fidelity estimation protocol explicitly asks for the demanding requirement of the classical estimation of the outcome probability $|\!\braket{\bar{x}_i|C_i|\psi}\!|^2$. Let us see how strong such a requirement is. Calling $p_{\epsilon_a}^{(i)}$ the classical estimation of $|\!\braket{\bar{x}_i|C_i|\psi}\!|^2$ within an additive error $\epsilon_{a}$, then $|k^{-1}\sum_{i}[(d+1)p_{\epsilon_a}^{(i)}-1]-\tilde{\mathcal{F}}|\le \epsilon_a(d+1)$, i.e., to ensure a small additive error on the estimation of the unbiased estimator in Eq.~\eqref{unbiasedshadow}, one should require $\epsilon_a\sim d^{-1}$, which rules out the sampling method used in Sec.~\ref{Sec: DFE} leading to an exponential scaling in $n$ for any state. Instead, if $q_{\epsilon_r}^{(i)}$ is the classical estimation of $|\!\braket{\bar{x}_i|C_i|\psi}\!|^2$ with a small relative error $\epsilon_r$, then $|k^{-1}\sum_{i}[(d+1)q_{\epsilon_r}^{(i)}-1]-\tilde{\mathcal{F}}|\le (\tilde{\mathcal{F}}+1)\epsilon_r\le 2\epsilon_r$ (almost surely). In other words, only if one is able to estimate the outcome probability within a small relative error is a shadow fidelity estimation protocol then possible. Note that the ability to compute all the outcome probabilities of $\ket{\psi}$, as well as the marginals, within a small relative error leads to the ability of sampling from the outcome distribution, see~\cite{bravyi2016ImprovedClassicalSimulation}, for $\psi$ and, therefore, quantum computation conducted by such states is entirely classical.

In what follows, we use the classical simulation method introduced in Sec.~\ref{Sec: SRE} that is able to estimate $|\!\braket{\bar{x}_i|C_i|\psi}\!|^2$ with no error. As explained below, this choice does not feature a loss of generality for the purpose of the paper.

Repeating steps $(i)$, $(ii)$ and $(iii)$ for $l$ times, and defining $\overline{\mathcal{F}}:=\operatorname{median}\{\mathcal{F}_{s}~|~s=1,\ldots, l\}$, we have the following theorem:

\emph{Theorem \hypertarget{thshadow}{3}}
Let $\ket{\psi}$ be a state prepared by a circuit containing a number $t$ of $T$ gates. Let $N_{\tilde{\psi}}$ be the number of times one needs to prepare $\tilde{\psi}$ on a quantum hardware, and let $N_{cl}$ be the number of classical resources for the classical estimation of the output probabilities. The number of resources necessary to estimate $\mathcal{F}(\ket{\psi},\tilde{\psi})$ within a error $\epsilon$, given by $\mathcal{N}_S=N_{\tilde{\rho}}\times N_{Cl}$, is
\be\label{eq:shadsam}
\mathcal{N}_S=\Theta(2^t\epsilon^{-2}\ln \delta^{-1}(n+t)^3)
\ee
In particular, the sample complexity is $N_{\tilde{\rho}}\equiv k\times l=\Theta(\epsilon^{-2}\ln \delta^{-1})$ where $l=8\ln 2\delta^{-1}$, while the number of classical resources $N_{cl}=\Theta(2^t(n+t)^{3})$.

\emph{Proof.}
The sample complexity bounded as $N_{\tilde{\rho}}\equiv k\times l=\Theta(\epsilon^{-2}\ln \delta^{-1})$ is to be found in~\cite{kliesch2021TheoryQuantumSystem}, while the scaling of the classical postprocessing complexity is derived in Sec.~\ref{Sec: strongsimulation} and then the total complexity $\mathcal{N}_S$ is given by the product; cf.Eq.~\ref{complexity}.

Let us make a couple of remarks about the theorem.

\emph{Remark 1}
One could argue about the existence of other simulation methods beyond the stabilizer formalism as matrix product decompositions (tensor network) or matchgates circuits. The question is can these methods provide better scalings for the resources in Theorem~III? The answer is ``no'' because the entire protocol relies on the extraction of a random Clifford circuit drawn uniformly at random according to the Haar measure over the Clifford group $\mathcal{C}(n)$ [see step $(i)$]; a state evolved by a random Clifford circuit $C$ is, with overwhelming probability, far beyond being easily encodable via tensor network decomposition; in other words, $C\ket{0^n}$ features a large bond dimension. On the other hand, a random Clifford circuit is far beyond being a match-gate circuit, because CX gates are not matchgates. We conclude that shadow fidelity estimation protocol implicitly requires a simulation method within the stabilizer formalism that, as shown in Sec.~\ref{Sec: strongsimulation}, scale exponentially in the number of non-Clifford gates in the circuit.

\emph{Remark 2}
As anticipated, we used the simulation method for Clifford+T circuits described in Sec.~\ref{Sec: SRE}, which is definitely not a \textit{state of the art} method. We opted for this pedagogical choice for the sake of simplicity. Indeed, the best-known simulations method approximates outcome probabilities within a small relative error gaining only a square root advantage with respect to the exponential scaling in $t$. The best known simulation algorithms are able to estimate $|\!\braket{\bar{x}_i|C_i|\psi}\!|^2$ within a relative error $\epsilon_r$ and a failure probability $p_f$ in time $\Theta(2^{\beta t}t^3\epsilon_{r}^{-2}\ln p_{f}^{-1})$, where $0<\beta\le 1/2$ (see~\cite{bravyi2016ImprovedClassicalSimulation}). In other words, we have no loss of generality in concluding that Theorem~\hyperlink{thshadow}{3} tells us that a shadow fidelity protocol is possible as long as the number of $T$ gates is $t=\mathcal{O}(\log_2 n)$, i.e., the same threshold that makes quantum computation by $\ket{\psi}$ entirely classical.

In the next section, we are going to make some comparisons between the above introduced protocols.

\subsection{Pauli fidelity estimation versus shadow fidelity estimation}\label{comparisonSFEDFE}
In this section, we compare the two protocols discussed in this paper and highlight their differences in resource utilization. It is important to note that both protocols are inefficient and this inefficiency is governed by nonstabilizerness. However, they differ in the way they use resources. One key difference is that the sampling complexity of shadow fidelity estimation is independent of the size of the Hilbert space, unlike Pauli fidelity estimation. However, it should be noted that in order to achieve efficient classical postprocessing, shadow fidelity estimation requires strong classical simulation for states to be certified. Additionally, it is worth noting that shadow fidelity estimation and Pauli fidelity estimation use different types of measurement data; while shadow fidelity estimation uses randomly selected bases for measurement, Pauli fidelity estimation uses the expectation values of observables. Correspondingly, they differ in their setups for experimental implementations. Let us conclude the section with the following remark. We recall that the (bound on) number of resources for the two protocols, to certify a given state $\psi\equiv \st{\psi}$, to ensure a small error and a high success probability are 
\ba
\mathcal{N}_P&=&\mathcal{O}(\exp M_{0}(\psi)), \quad \text{Pauli fidelity estimation}\,,\nonumber\\
\mathcal{N}_S&=&\mathcal{O}(\exp t), \quad \text{shadow fidelity estimation}\,,
\ea

where we neglected the dependency from $\epsilon$ and $\delta$ displayed in Eqs.~\eqref{eq:pausam} and \eqref{eq:shadsam} respectively. By a closer look at the scaling of the resources, one realizes that, for generic states, Pauli fidelity estimation performs better than shadow fidelity estimation. Indeed, as shown in Sec.~\ref{Sec: SRE}, one has $M_{0}(\psi)\le \nu(\psi)$, for $\nu(\psi)$ being the stabilizer nullity, and thus
\be
M_{0}(\psi)\le t\,,
\ee
for $t$ being the number of non-Clifford gates used to prepare $\psi$. Note that the above inequality becomes strict in many cases. One trivial example is provided by the sequential application of $T$ gates, because $T^2=S$: while the zero-stabilizer entropy does not change, being invariant under Clifford operations, the number of $T$ gates does; only using a compilation procedure aimed to reduce the \textit{T count}, as the one employed in~\cite{heyfron2018EfficientQuantumCompiler}, one can get rid of additional useless $T$ gates.

Let us conclude the paper with the extensions of the above results to the certification of quantum processes.
\subsection{Quantum processes}
In this section, we show that the stabilizer R\'enyi entropy of the Choi state $\ket{U}$ bounds the resources needed to perform a quantum process verification. Suppose one wants to characterize the quality of the application of a given unitary operator $U$. This task occurs in many quantum algorithms and the quantum Fourier transform provides a nice example. Let $\mathcal{U}$ be the quantum map (in general non unitary) applied by the quantum processor. One way to certify the quality of $\mathcal{U}$ is through the average gate fidelity\cite{roth2018RecoveringQuantumGates,kliesch2021TheoryQuantumSystem,flammia2011DirectFidelityEstimation,reich2013OptimalStrategiesEstimating}:
\be
F_{avg}(U):=\int \de\psi\, \mathcal{F}(U\ket{\psi},\mathcal{U}(\psi))
\ee
i.e., the average fidelity between the application of the target unitary on $\ket{\psi}$ and the quantum map on $\psi\equiv\st{\psi}$, according to the Haar measure $\de \psi$. One can easily show, via a Kraus operator expansion, see Appendix~\ref{app:unitaryoperators}, that $F_{avg}(U)=\mathcal{F}_{U}+\mathcal{O}(d^{-1})$, where
\be
\mathcal{F}_{U}:=\frac{1}{d^{4}}\sum_{\mu\nu}\tr(P_{\mu}UP_{\nu}U^{\dag})\tr(P_{\mu}\mathcal{U}(P_{\nu}))
\ee
is the {\em entanglement fidelity} between $U$ and the quantum map $\mathcal{U}(\cdot)$\cite{kliesch2021TheoryQuantumSystem}.  Let us use the same trick as before: define a probability distribution $\Xi_{U}:=\{\tr^{2}(P_{\mu}UP_{\nu}U^{\dag})/d^4\,|\, \mu,\nu=1,\ldots,d^2\}$, and rewrite $\mathcal{F}_{U}$ as the average of $\mathcal{X}_{\mu\nu}:=\tr(P_{\mu}\mathcal{U}(P_{\nu}))/\tr(P_{\mu}UP_{\nu}U^{\dag})$ on the probability distribution $\Xi_{U}$, i.e., $\mathcal{F}_{U}=\aver{\mathcal{X}_{\mu\nu}}_{\Xi_U}$.
$\mathcal{F}_{U}$ can thus be estimated via Monte Carlo methods by sampling $k$ pairs of Pauli operators $(P_{1},P_{1}^{\prime}),\dots, (P_{k},P_{k}^{\prime})$ according to the probability distribution $\Xi_{U}$. We quantify the resources as the number of uses $N_{\mathcal{U}}$ of the channel $\mathcal{U}$. Note that, from Lemma \hyperlink{choistatelemma}{1}, the probability distribution $\Xi_{U}$ coincides with the probability distribution associated with the Choi state $\ket{U}$. The following theorem provides bounds for the number of resources needed to estimate $\mathcal{F}_{\mathcal{U}}$ in terms of the stabilizer R\'enyi entropy $M_{\alpha}(\ket{U})$.

\emph{Theorem \hypertarget{theorem2}{4}}
The number of resources $N_{\mathcal{U}}$ to estimate $\mathcal{F}_{\mathcal{U}}$ with accuracy $\epsilon$ and success probability $1-\delta$ is bounded by
\be
\frac{2}{\epsilon^{2}}\ln(2/\delta)\exp[M_{2}(\ket{U})]\le N_{\mathcal{U}}\le \frac{64}{\epsilon^{4}}\ln(2/\delta)\exp[M_{0}(\ket{U})]
\ee

See Appendix~\ref{app:theorem2} for the proof. We found that the nonstabilizerness of the Choi state $\ket{U}$ is a direct quantifier of the hardness in verifying the correct application of a target unitary $U$. The result presented in Lemma \hyperlink{choiotoclemma}{2} tells the meaning of the stabilizer R\'enyi entropy $M_{\alpha}(\ket{U})$:

\emph{Corollay 4}
The resources $N_{\mathcal{U}}$ are bounded:
\be
\Theta(OTOC_{8}(U)^{-1})\le N_{\mathcal{U}}\le \Theta(\exp[\nu(U)])
\ee
where $OTOC_{8}(U)$ is defined in Lemma \hyperlink{choiotoclemma}{2} and $\nu(U)$ is the unitary stabilizer nullity.
Therefore, the more chaotic the evolution generated by $U$ is, the smaller the out-of-time-order correlators are~\cite{bravyi2019SimulationQuantumCircuits} and the harder the certification via direct fidelity estimation is. 

In the same fashion of doped stabilizer states, the doped Clifford circuits provide an important class of circuits to look at. A $t$-doped Clifford circuit~\cite{leone2021QuantumChaosQuantum,haferkamp2020QuantumHomeopathyWorks} consists of global layers of Clifford gates interleaved by single qubit $T$ gates. Bravyi and Gosset\cite{bravyi2016ImprovedClassicalSimulation} proved that the classical simulations of such circuits scale exponentially with the number of $T$ gates, while in \cite{leone2021QuantumChaosQuantum} we proved that to mimic quantum chaotic evolution, a quantum circuit should contain at least $\mathcal{O}(n)$ $T$ gates, showing the impossibility to simulate quantum chaos classically. In this scenario, we ask the question of whether quantum chaos can be effectively certified by the above fidelity estimation protocol. The following theorem determines the scaling of $N_{\mathcal{U}}$ with the number $t$ of $T$ gates.

\emph{Corollary \hypertarget{dopedCliffordresources}{5}}
Let $\aver{N_{\mathcal{C}_t}}$ be the average number of resources to verify a $t$-doped Clifford circuit $C_{t}$; then it increases exponentially with $t$:
\be
\Theta(\exp[t\log_24/3])\le\aver{N_{\mathcal{C}_t}}\le \Theta(\exp[t])\le \Theta(d^2)
\label{ultimaeq}
\ee
The answer is no: quantum chaos cannot be efficiently certified, as the protocol is efficient up to $\mathcal{O}(\log_2 n)$ $T$ gates injected in a Clifford circuit. See Appendix~\ref{app:dopedCliffordresources} for the proof. Let us now briefly comment on the scalings of the bounds in Eq. \eqref{ultimaeq}. First, note that such scalings are the same as those in Corollary \hyperlink{averagestabresources}{1}: while for states the resources are upper bounded by $\Theta(d)$ and the bound is saturated after the injection of ``only'' $n$ non-Clifford gates, for unitary operators the injection of more than $n$ non-Clifford resources makes the verification even harder.
%%%%%%%%%%%%%%%%%%%%%%%%%%%%%%%%%%%%%%%%%%%%%%%%%%%%%%%%%%%%%%%%%%%%%%%%%%%%%%%%%%%%%%%%%%%%%%%%%%%%%%%%%%%%%%%%%%%%%%%%%%%%%%%%%%%%%%%%%%%%%%%%%%%%%%%%%%%%%%%%%%%%%%%%%%%%%%%%%%%%%%%%%%%%%%%%%%%%%%%%%%%%%%%%%%%%%%%%%%%%%%%%%%%%%%%%%%%%%%%%%%%%%%%%%%%%%%%%%%%%%%%%%%%%%%%%%%%%%%%%%%%%%%%%%%%%%%%%%%%%%%%%%%%%

\section{Summary and discussion} 
In this paper, we showed the tight connection underlying quantum certification via direct fidelity estimation, nonstabilizerness, and chaos. We showed that the complexity of Pauli fidelity estimation and shadow fidelity estimation  scales exponentially with the number of non-Clifford gates and thus with the nonstabilizerness $M(\ket{\psi})$. This fact implies the impossibility of such protocols to certify all the states $\ket{\psi}$ beyond the efficiency threshold $M(\ket{\psi})=\mathcal{O}(\log_2 n)$. Remarkably, the protocol fails to certify all those states which turn out to be useful to achieve quantum speedups. In other words, there is no free lunch: any quantum certification protocol aimed to directly estimate the fidelity between the theoretical state and the actual state becomes inefficient, and this inefficiency is governed by nonstabilizerness, the resource which makes quantum technology truly quantum. However, we note that the inefficiency is due to the wide applicability of such protocols: there exist other protocols aimed to certify particular sets of states that, although possessing a high amount of nonstabilizerness, can be efficiently certified. One prominent example is the set of hypergraph states. Such states feature an extensive amount of nonstabilizerness, making both Pauli fidelity estimation and shadow fidelity estimation unfeasible, but possess efficiently encodable stabilizer operators that make their certification possible, as shown in Sec.~\ref{Sec: DFE} (see Corollary~\hyperlink{cor: gDFe}{3}). The scope of this work is to rule out the use of general and widely applicable protocols for direct fidelity estimation and, at the same time, to leave the window open for state-aware protocols aimed to certify certain specific classes of states. After all, the class of quantum states that are truly useful for a quantum computational speed up is of measure zero in the Hilbert space~\cite{gross2009MostQuantumStates} and, therefore, there is no need for a general protocol able to certify every quantum state.

{\em Acknowledgments}---L.L. and S.F.E.O. thank Sarah True for helpful comments. The authors acknowledge support from NSF Award No. 2014000. The work of L.L. and S.F.E.O. was supported in part by College of Science and Mathematics Dean’s Doctoral Research Fellowship through fellowship support from Oracle, Project ID R20000000025727. A.H. acknowledges financial support from PNRR MUR project PE0000023-NQSTI and PNRR MUR project CN\_00000013 -ICSC. 

%one has to estimate the outcome probabilities in a computational basis state of the theoretical state $U\ket{\psi}$ for each $U\sim \mu_{Cl}$

\appendix
\onecolumngrid
\section*{Appendix}

%%%%%%%%%%%%%%%%%%%%%%%%%%%%%%%%%%%%%%%%%%%%%

\section{Stabilizer R\'enyi entropy}
\subsection*{Proof of Lemma \hypers{choistatelemma}{1}}\label{app:choistatelemma}

In this section, we prove that the nonstabilizerness of the Choi–Jamiołkowski isomorphism can be measured as the R\'enyi entropy of the probability distribution $\Xi_{U}$ whose elements are
\be
\Xi_{U}(P,P^{\prime})=\frac{\tr^2(PUP^{\prime}U^{\dag})}{d^4}
\ee
Recall that the Choi isomorphism is a map from the space of operator $\mathcal{B}(\mathcal{H})$ to state vectors in $\mathcal{H}^{\otimes 2}$. Let $U$ be a unitary operator; its Choi isomorphism $\ket{U}\in\mathcal{H}^{\otimes 2}$ is defined as
\be
\ket{U}:=(\bbbone\otimes U)\ket{I}, \quad \ket{I}=\frac{1}{\sqrt{d}}\sum_{i}\ket{i}\otimes\ket{i}
\ee
Let us compute the stabilizer R\'enyi entropy of $\ket{U}$. Since now we are working with states of $\mathcal{H}^{\otimes 2}$, the Pauli group is $\mathbb{P}(2n)=\mathbb{P}(n)\otimes\mathbb{P}(n)$ and the coefficients of the probability distribution for $\ket{U}$ read
\be
\Xi_{U}(P\otimes P^{\prime})=\frac{1}{d^2}\tr^{2}(P\otimes P^{\prime}\st{U}), \quad P,P^{\prime}\in\mathbb{P}(n)
\ee
the stabilizer R\'enyi entropy reads:
\be
M_{\alpha}(\ket{U})=\frac{1}{1-\alpha}\log_2 \sum_{P,P^{\prime}}|\Xi_{U}(P\otimes P^{\prime})|^{\alpha}-2\log_2 d
\ee
Let us prove that the coefficients $\tr(P\otimes P^{\prime}\st{U})\propto \frac{1}{d}\tr(PUP^{\prime}U^{\dag})$ up to a global phase $\pm 1$. First, it is well-known that the trace is invariant under partial transpose: let $A,B\in\mathcal{B}(\mathcal{H})$ two operators on $\mathcal{H}$, then the partial transpose is defined as $(A\otimes B)^{T_2}:=A\otimes B^{T}$, where $B^{T}$ is the transpose of $B$.
\be
\tr(P\otimes P^{\prime}\st{U})=\tr(P\otimes P^{\prime}\st{U})^{T_{2}}=\pm\tr(P\otimes P^{\prime}\st{U}^{T_{2}})
\ee
where the $\pm$ comes from the fact that $P^{\prime T}\propto P^{\prime}$ up to a sign (because $Y^{T}=-Y$, $X^T=X$, and $Z^T=Z$). Now
\be
\st{U}^{T_2}=(\bbbone\otimes U^{T})\st{I}^{T_2}(\bbbone\otimes U^{*})=(\bbbone\otimes U^{T})\frac{\hat{S}}{d}(\bbbone\otimes U^{*})=\frac{\hat{S}}{d}(U^{T}\otimes U^{*})
\ee
where $\hat{S}$ is the swap operator. The fact that $\st{I}^{T_2}=\frac{\hat{S}}{d}$ can be checked straightforwardly, then
\be
\tr(P\otimes P^{\prime}\st{U})=\frac{\pm1}{d}\tr(PU^{T} P^{\prime}U^{*})=\frac{\pm1}{d}\tr(  P^{\prime}UPU^{\dag})
\ee
Thus we obtain that the elements of the probability distribution $\Xi_{U}$ read
\be
\Xi_{U}(P,P^{\prime})=\frac{1}{d^2}\tr^{2}(P\otimes P^{\prime}\st{U})=\frac{1}{d^4}\tr^{2}(P^{\prime}UPU^{\dag})
\ee
and the lemma follows straightforwardly.

\subsection*{Proof of Lemma \hypers{choiotoclemma}{2}}\label{app:choiotoclemma}
From Lemma \hyperlink{choistatelemma}{1}, we have that:
\be
M_{\alpha}(\ket{U})=\frac{1}{1-\alpha}\log_2\sum_{P,P^\prime}\frac{\tr^{2\alpha}(PUP^\prime U^{\dag})}{d^{4\alpha}}-\log_2 d^2=\frac{1}{1-\alpha}\log_2\frac{1}{d^2}\sum_{P,P^\prime}\frac{\tr^{2\alpha}(PUP^\prime U^{\dag})}{d^{2\alpha}}
\ee
To prove the Lemma, we recall the following identity:
\be
\hat{S}=\frac{1}{d}\sum_{P}P^{\otimes2}
\ee
where $\hat{S}$ is the swap operator. Note that:
\be
\frac{\tr(PUP^\prime U^{\dag})}{d}\frac{\tr(PUP^\prime U^{\dag})}{d}=\frac{1}{d^2}\sum_{P_1}\frac{\tr(PUP^\prime U^{\dag}P_1PUP^\prime U^{\dag}P_1)}{d}\equiv \frac{\tr(\langle PUP^\prime U^{\dag}P_1PUP^\prime U^{\dag}P_1\rangle_{P_1} )}{d}
\ee
We thus can recursively use the above identity and arrive to
\ba
\frac{\tr^{2\alpha}(PUP^\prime U^{\dag})}{d^{2\alpha}}=d^{-1}\tr[\langle P_{2\alpha}\prod_{i=1}^{2\alpha}UPU^{\dag}P^{\prime}P_{i-1}P_{i}\rangle ]=otoc_{2\alpha}(P,P^\prime)
\ea
where $\aver{\cdot}\equiv d^{-2}\sum_{P_i\in\mathbb{P}(n)}$ for all $i=1,\dots,2\alpha$, while $P_{0}\equiv\bbbone$. Let us write the above explicitly for $\alpha=2$:
\ba
\frac{\tr^{4}(PUP^\prime U^{\dag})}{d^{4}}&=&d^{-1}\tr(\langle P_{4}P^{(U)}P^\prime P_1P^{(U)}P^\prime P_1P_2P^{(U)}P^\prime P_2P_3P^{(U)}P^\prime P_3P_{4}\rangle_{P_1,\ldots,P_4} )\nonumber\\&=&d^{-1}\tr(\langle P^{(U)}P^\prime P_1P^{(U)}P^\prime P_1P_2P^{(U)}P^\prime P_2P_3P^{(U)}P^\prime P_3\rangle_{P_1,\ldots,P_4} )=otoc_{8}(P,P^{\prime})
\ea
Note that the above holds for any integer $\alpha>1$.

\section{Quantum states certification}
\subsection{Proof of Theorem \hypers{theorempurestates}{2}}\label{app:theorempurestates}
In this section, we give proof of the main theorem in the manuscript. Some parts of the proof are inspired by the work of Flammia et. al \cite{flammia2011DirectFidelityEstimation}, see also \cite{kliesch2021TheoryQuantumSystem}. We prove the two bounds separately.
\begin{enumerate}[label=(\roman*)]
\item {\bf Lower bound:}
Here we need to lower bound the necessary resources such that the estimator $\tilde{\mathcal{F}}=\frac{1}{k}\sum_{P\in\mathbb{K}}\tilde{X}_{P}$, defined in the main text, obeys to
\be
\operatorname{Pr}(|\mathcal{F}-\tilde{\mathcal{F}}|\le \epsilon)\ge1-\delta
\ee
To prove it, define $m:=\min_{P}|\tr(P\psi)|$ and note that $|\tilde{X}_{P}|\le m^{-1}$. Using Hoeffding's inequality\cite{hoeffding1963ProbabilityInequalitiesSums}, one can bound the probability
\be
\operatorname{Pr}(|\mathcal{F}-\tilde{\mathcal{F}}|\le \epsilon)\ge1- 2\exp\left[-\frac{k\epsilon^2}{2m^{-2}}\right]\,.
\ee
To have the probability lower bounded by $1- \delta$, the number of samples $k$ must be:
\be
k=\frac{2}{\epsilon^2m^2}\ln(2/\delta)\,.
\ee
Setting the number of copies $c_{P}(\tilde{\psi})$ of the state $\tilde{\psi}$ to determine each sampled $P$ to be one (one-shot measurements), i.e., $c_{P}(\tilde{\psi})=1$ $\forall P$, one has that $N_{\tilde{\psi}}\equiv k$. Let us lower bound the number of resources $N_{\tilde{\psi}}$. Let $P\in\mathbb{P}(n)$, then the average of $|\tr(P\psi)|$ over the state dependent probability distribution $\Xi_{\psi}$ is upper bounded by
\be
\aver{|\tr(P\psi)|}_{\Xi_{\psi}}\le \sqrt{\aver{\tr^{2}(P\psi)}_{\Xi_{\psi}}}=\sqrt{\exp[-M_{2}(\ket{\psi})]}  \,.
\ee
Then since $m=\min_{P}|\tr(P\psi)|$, one trivially has $m\le \aver{|\tr(P\psi)|}_{\Xi_{\psi}}$ and thus $m\le \sqrt{\exp[-M_{2}(\ket{\psi})]}$. Thus, the number of resources $N_{\tilde{\psi}}$ is lower bounded
\be
N_{\tilde{\psi}}\ge \frac{2}{\epsilon^{2}}\ln(2/\delta)\exp[M_{2}(\ket{\psi})]\,.
\ee

    \item {\bf Upper bound:}
    
Let $\psi=\st{\psi}$ be the state we want to verify. Let us define the following operator (in the Pauli basis fashion):
\be
\tr(P\psi_{cut}):=\begin{cases}\tr(P\psi), \quad \text{if}\, |\tr(P\psi)|\ge \frac{\epsilon}{2\sqrt{2}}\sqrt{\exp[-M_{0}(\psi)]}\\
0, \quad \text{otherwise}
\end{cases}
\ee
and its normalized version $\psi^{\prime}:=\psi_{cut}/\norm{\psi_{cut}}_{2}$. Define $\mathbb{Q}:=\{P\in\mathbb{P}(n)\,|\,|\tr(P\psi)|\ge \epsilon/2/\sqrt{2}\sqrt{\exp[-M_{0}(\psi)]}\}$ so that $\psi^\prime$ in the Pauli basis reads:
\be
\psi^{\prime}=\frac{1}{\sqrt{\frac{1}{d}\sum_{P\in\mathbb{Q}}\tr^{2}(P\psi)}}\frac{1}{d}\sum_{P\in\mathbb{Q}}\tr(P\psi)P\,.
\ee
Let us evaluate the difference between $\mathcal{F}^{\prime}(\psi^{\prime},\tilde{\psi}):=\tr(\psi^{\prime}\tilde{\psi})$ and the true fidelity $\mathcal{F}(\ket{\psi},\tilde{\psi})$:
\be
|\mathcal{F}^{\prime}-\mathcal{F}|\le \norm{\psi^\prime-\psi}_{2}=\sqrt{2(1-\tr(\psi\psi^{\prime}))}
\ee
In the above we used $\tr(\psi^{\prime2})=1$. Let us evaluate $\tr(\psi\psi^\prime)$ by writing it in the Pauli basis:
\be
\tr(\psi\psi^\prime)=\frac{1}{\norm{\psi_{cut}}_{2}}\frac{1}{d}\sum_{P\in\mathbb{Q}}\tr^{2}(P\psi)=\sqrt{\frac{1}{d}\sum_{P\in\mathbb{Q}}\tr^{2}(P\psi)}=\sqrt{1-\frac{1}{d}\sum_{P\in\bar{\mathbb{Q}}}\tr^{2}(P\psi)}\ge\sqrt{ 1-\frac{\epsilon^2\exp[-M_{0}(\psi)]|\mathbb{Q}|}{8d}}\,,
\ee
where $\bar{\mathbb{Q}}$ is the complement set of $\mathbb{Q}$. Note that $|\bar{\mathbb{Q}}|=\card(\psi)-|\mathbb{Q}|$---where $\card(\psi):=|\{P\,|\,\tr(P\psi)\neq 0\}|$---and that $\card(\psi)/d=\exp[M_{0}(\psi)]$. We obtain $\tr(\psi\psi^\prime)\ge \sqrt{1-\epsilon^2/8}\ge 1-\epsilon^2/8$ and thus:
\be
|\mathcal{F}^{\prime}-\mathcal{F}|\le \frac{\epsilon}{2}\,.
\ee
Note that $\mathcal{F}^{\prime}$ can be estimated in the same fashion as $\mathcal{F}$:
\be
\mathcal{F}^{\prime}=\frac{1}{d}\sum_{P}\tr(P\tilde{\psi})\tr(P\psi^{\prime})=\aver{X_{P}^{\prime}}_{\Xi_{\psi^\prime}}\,,
\ee
where the average is taken over the probability distribution $\Xi_{\psi^\prime}$ whose elements are
\be
\Xi_{\psi^\prime}(P)=\begin{cases}\frac{\tr^{2}(\psi P)}{\sum_{P\in\mathbb{Q}}\tr^2(P\psi)}, \quad P\in\mathbb{Q}\\0,\quad \text{otherwise}
\end{cases}
\ee
and $X_{P}^{\prime}:=\frac{\tr(P\tilde{\psi})}{\tr(P\psi^{\prime})}$. Thus, we define $\tilde{\mathcal{F}^\prime}$ the estimator of $\mathcal{F}^{\prime}$ obtained by sampling the probability distribution $\Xi_{\psi^\prime}$ and by experimentally measuring $X_{P}^{\prime}\in\mathbb{K}^{\prime}$:
\be
\tilde{\mathcal{F}}^{\prime}=\frac{1}{k}\sum_{P\in\mathbb{K}^{\prime}}\tilde{X}_{P}^{\prime}\,,
\ee
where $\tilde{X}_{P}^{\prime}=\frac{1}{\tr(\psi^{\prime} P)}\frac{1}{c_{P}(\tilde{\psi})}\sum_{j=1}^{c_{P}(\tilde{\psi})}\mathcal{P}_{Pj}(\tilde{\psi})$, $c_{P}(\tilde{\psi})$ is the number of copies $\tilde{\psi}$ used to estimate $P\in\mathbb{K}$ and $\mathcal{P}_{Pj}(\tilde{\psi})$ the outcome of a one-shot measurement of $P$. 

Let us prove that, setting $N_{\tilde{\psi}}\le \frac{64}{\epsilon^{4}}\ln(2/\delta)\exp[M_{0}(\psi)]$, we have
\be
\operatorname{Pr}(|\mathcal{F}-\tilde{\mathcal{F}^{\prime}}|\le \epsilon)\ge 1-\delta\,.
\ee
First
\be
|\mathcal{F}-\tilde{\mathcal{F}}^{\prime}|\le |\mathcal{F}-{\mathcal{F}}^{\prime}|+|\mathcal{F}^{\prime}-\tilde{\mathcal{F}}^{\prime}|\le\frac{\epsilon}{2}+|\mathcal{F}^{\prime}-\tilde{\mathcal{F}}^{\prime}|\,.
\ee
Then note that $\operatorname{Pr}(|\mathcal{F}-\tilde{\mathcal{F}^{\prime}}|\le \epsilon)=\operatorname{Pr}(|\mathcal{F}^{\prime}-\tilde{\mathcal{F}^{\prime}}|\le \epsilon/2)$. Since $\mathbb{E}(\tilde{\mathcal{F}}^{\prime})=\mathcal{F}^{\prime}$, i.e., $\tilde{\mathcal{F}}^{\prime}$ is an unbiased estimator for $\mathcal{F}^{\prime}$, we can use Hoeffding's inequality:
\be
\operatorname{Pr}(|\mathcal{F}^{\prime}-\tilde{\mathcal{F}^{\prime}}|\le \epsilon/2)=1-2\exp\left[\frac{km^{\prime2}\epsilon^2}{8}\right]\,,
\ee
where $m^{\prime}:=\min_{P}|\tr(\psi^\prime P)|$ and thus $|\tilde{X}_{P}^{\prime}|\le m^{-1}$. To have that the probability is lower bounded by $1-\delta$, we impose that $c_{P}(\tilde{\psi})=1$ for any $P\in\mathbb{Q}$ and:
\be
N_{\tilde{\psi}}=k=\frac{8}{\epsilon^{2}m^{\prime2}}\ln(2/\delta)\,.
\ee
To prove the upper bound to the number of resources $N_{\tilde{\psi}}$ is sufficient to note that:
\be
m^{\prime}=\min_{P\in\mathbb{Q}}\frac{|\tr(P\psi)|}{\sqrt{\frac{1}{d}\sum_{P\in\mathbb{Q}}\tr^{2}(P\psi)}}\ge \min_{P\in\mathbb{Q}}|\tr(P\psi)|\ge \frac{\epsilon}{2\sqrt{2}}\sqrt{\exp[-M_{0}(\psi)]}\,,
\ee
where we exploited once again the fact that $\sqrt{\frac{1}{d}\sum_{P\in\mathbb{Q}}\tr^{2}(P\psi)}\le 1$. We finally obtain:
\be
N_{\tilde{\psi}}\le \frac{64}{\epsilon^{4}}\ln(2/\delta)\exp[M_{0}(\psi)]\,,
\ee
which concludes the proof.
\end{enumerate}
\subsection{Proof of Corollary \hypers{averagestabresources}{1}}\label{app:averagestabresources}
From the main theorem, we have that:
\be
\frac{2}{\epsilon^{2}}\ln(2/\delta)\aver{\exp[M_{2}(\psi_t)]}\le \aver{N_{\tilde{\psi}_t}}\le \frac{64}{\epsilon^{4}}\ln(2/\delta)\aver{\exp[M_{0}(\psi_t)]}\,.
\ee
The average of the left-hand side for states can be lower bounded through the Jensen inequality and we obtain:
\be 
\aver{N_{\tilde{\psi}_t}}\ge \frac{2}{\epsilon^{2}}\ln(2/\delta)\aver{\exp[M_{2}(\psi_t)]} \ge \frac{2}{\epsilon^{2}}\ln(2/\delta)\frac{1}{\aver{\tr{(Q\psi_{t}^{\otimes 4}})}}\,,
\ee 
where $Q=\frac{1}{d^{2}}\sum_{P\in\mathcal{P}_{n}}P^{\otimes 4}$. Then the average over $t$-doped stabilizer states $\ket{\psi_t}$ can be computed using the techniques in\cite{leone2022StabilizerRenyiEntropy,oliviero2021TransitionsEntanglementComplexity}. The result is shown in Eq. (13) of \cite{leone2022StabilizerRenyiEntropy}:
\be
\aver{\exp[M_{2}(\psi_t)]}\ge \frac{d+3}{4+(d-1)f_{+}^{t}}= \Theta(\exp[t\log_2 4/3])\,,
\ee
where $f_{+}=\frac{3d^2-3d -4}{4 (d^2 -1)}$ and this concludes the proof. 

The right-hand-side can be upper bounded using the stabilizer nullity. Recall that
\be
\exp[M_{0}(\psi_t)]\le \exp[\nu(\ket{\psi_{t}})]\,,
\ee
where $\nu(\ket{\psi_t})$ is the stabilizer nullity of the $t$-doped stabilizer state. We can write such a state as $\ket{\psi_t}=C_{t}\ket{0}^{\otimes n}$ where $C_{t}$ is a doped Clifford circuit, i.e., layers of Clifford operators interleaved by the action of single qubit $T$ gates. Then\cite{jiang2021LowerBoundTcount} we have the following chain of inequality:
\be
\nu(\ket{\psi_t})=\nu(C_{t}\ket{0}^{\otimes n})\le \nu(C_{t})\le t\,,
\ee
where $\nu(C_{t})$ is the unitary stabilizer nullity, introduced in\cite{jiang2021LowerBoundTcount}, which lower bounds the number of non-Clifford resources injected in a Clifford unitary operator. Therefore we obtain
\be
\exp[M_{0}(\psi_t)]\le \exp[t]\,.
\ee
Lastly note that since $M_{0}(\psi_t)\le \log_2 d$, we have $\aver{N_{\psi_t}}\le d$.

\section{Unitary operators}\label{app:unitaryoperators}
\subsection*{Entanglement fidelity}
In this section we prove that $F_{avg}=\mathcal{F}_{U}+\mathcal{O}(d^{-1})$, i.e., the average gate fidelity $F_{avg}$ is the entanglement fidelity $\mathcal{F}_{U}$ up to an error scaling as $\mathcal{O}(d^{-1})$. Let us start with the definition of average gate fidelity given in the main text:
\be
F_{avg}(U):=\int\de\psi\tr(U\psi U^{\dag}\mathcal{U}(\psi))\,.
\ee
By expanding $\mathcal{U}$ in terms of Kraus operator $A_{\alpha}$ one can write the above as
\be
F_{avg}=\sum_{\alpha}\int\de\psi\tr(U\psi U^{\dag}A_{\alpha}\psi A_{\alpha}^{\dag})\,.
\ee
By the well-known identity\cite{page1993AverageEntropySubsystem} $\int\de\psi\psi^{\otimes 2}=[d(d+1)]^{-1}(\bbbone+\hat{S})$, one has
\be
F_{avg}=\frac{\frac{1}{d}\sum_{\alpha}\tr(U^{\dag}\otimes UA_{\alpha}\otimes A_{\alpha}^{\dag})+1}{d+1}\,.
\ee
Multiplying by $\bbbone^{\otimes 2}\equiv \hat{S}\hat{S}$ and by expanding both $\hat{S}U\otimes U^{\dag}$ and $\hat{S}A_{i}\otimes A_{i}^{\dag}$ in terms of the Pauli operators on $\mathcal{H}^{\otimes 2}$, we have
\be
\tr(U^{\dag}\otimes UA_{\alpha}\otimes A_{\alpha}^{\dag})=\frac{1}{d^2}\sum_{\mu\nu}\tr(P_{\mu}UP_{\nu}U^{\dag})\tr(P_{\mu}A_{\alpha}P_{\nu}A_{\alpha}^{\dag})\,.
\ee
Finally $F_{avg}=\mathcal{F}_{U}+\mathcal{O}(d^{-1})$, where we defined
\be
\mathcal{F}_{U}:=\frac{1}{d^{4}}\sum_{\mu\nu}\tr(P_{\mu}UP_{\nu}U^{\dag})\tr(P_{\mu}\mathcal{U}(P_{\nu})).
\ee

\subsection*{Proof of Theorem \hypers{theorem2}{4}}\label{app:theorem2}
In this section, we prove Theorem \hyperlink{theorem2}{4}. We give the proof for the lower and the upper bound separately.
\begin{enumerate}[label=(\roman*)]
\item {\bf Lower bound:}
    
Let $(P_{1},P_{1}^{\prime}),\ldots, (P_{k},P_{k}^{\prime})$ be $k$ pairs of Pauli operators sampled at random according to the probability distribution $\Xi_U$ and labeled by $i=1,\ldots,k$. Let $\tilde{\mathcal{F}}_{\mathcal{U}}=\frac{1}{k}\sum_{i=1}^{k}\tilde{\mathcal{X}}_{i}$ be an estimator for $\mathcal{F}_{U}$, i.e., $\mathbb{E}[\tilde{\mathcal{F}}_{\mathcal{U}}]=\mathcal{F}_{\mathcal{U}}$, where 
\be
\mathcal{X}_{i}=\frac{1}{\tr(UP_{i}U^{\dag}P_{i}^{\prime})}\frac{1}{c_{i}(\mathcal{U})}\sum_{j=1}^{c_{i}(\mathcal{U})}\mathcal{P}_{ij}(\mathcal{U})
\ee
where $\mathcal{P}_{ij}(\mathcal{U})$ is the $j$th measurement of $\tr(\mathcal{U}(P_{i})P_{i}^{\prime})$ and $c_{i}(\mathcal{U})$ are the number of copies needed to estimate a given pair $(P_{i},P^{\prime}_{i})$ for $i=1,\ldots, k$. Following the proof of Theorem \hyperlink{theorempurestates}{2}, define $m_U:=\min_{P,P^{\prime}}|\tr(U^{\dag}PUP^{\prime})|/d$ and note that $|\tilde{X}_{i}|\le m_{U}^{-1}$. Using Hoeffding's inequality we have that
\be
\operatorname{Pr}(|\tilde{\mathcal{F}_{\mathcal{U}}}-\mathcal{F}_{\mathcal{U}}|\le\epsilon)\ge 1-2\exp\left[-\frac{k\epsilon^2}{2m_{U}^{-2}}\right]\,.
\ee
Thus, by imposing the probability to be lower bounded by $1-\delta$ and by setting $c_{i}(\mathcal{U})=1$ for any $i$ (i.e., one-shot measurements) one gets
\be
\mathcal{N}_{\mathcal{U}}=\frac{2}{\epsilon^2m_{U}^2}\ln(2/\delta)
\ee
to prove the lower bound is sufficient to note that
\be
m_{U}:=\min_{P,P^{\prime}}|\tr(U^{\dag}PUP^{\prime})|/d\le \aver{d^{-1}|\tr(U^{\dag}PUP^{\prime})|}_{\Xi_{U}}\le d^{-1}\sqrt{\sum_{P,P^{\prime}}\frac{\tr^4(U^{\dag}PUP^{\prime})}{d^4}}\equiv \sqrt{\exp[-M_{2}(\ket{U})]}\,,
\ee
where $M_{2}(\ket{U})$ is the stabilizer R\'enyi entropy of the Choi state $\ket{U}$; cf. Lemma \hyperlink{choistatelemma}{1}.
\item {\bf Upper bound:}

To prove the upper bound let us define an auxiliary operator $U_{cut}$ with a similar technique of the one used for pure states. Define the following coefficients:
\be
\tr(U_{cut}^{\dag}PU_{cut}P^{\prime}):=\begin{cases} \tr(U^{\dag}PUP^{\prime}), \quad \text{if}\, |\tr(U^{\dag}PUP^{\prime})|/d\ge  \theta \sqrt{\exp[-M_{0}(\ket{U})]}\\
0, \quad \text{otherwise}
\end{cases}
\label{Ucutdef}
\ee
and $\mathbb{Q}_{U}:=\{P,P^{\prime}\,|\, \tr(U_{cut}^{\dag}PU_{cut}P^{\prime})\neq 0\}$. Now define the operator $U^{\prime}$ such that:
\be
 \hat{S}U^{\prime\dag}\otimes U^{\prime}=\frac{1}{\sqrt{\sum_{P,P^\prime\in\mathbb{Q}_{U}}\tr^2(U^\dag PUP^\prime)}}\sum_{P,P^\prime\in\mathbb{Q}_{U}}\tr(U^\dag PUP^\prime)P\otimes P^\prime\,.
\ee
Let us evaluate the difference between $\mathcal{F}_{U^{\prime}}:=\frac{1}{d^2}\sum_{\alpha}\tr(U^{\prime\dag}\otimes U^{\prime}A_{\alpha}\otimes A_{\alpha}^{\dag})$ and $\mathcal{F}_{U}$ defined in the main text:
\be
|\mathcal{F}_{U^\prime}-\mathcal{F}_{U}|\le \frac{1}{d^2}\norm{\sum_{\alpha}A_{\alpha}\otimes A_{\alpha}^{\dag}}_{2}\norm{U^{\prime\dag}\otimes U^{\prime}-U^{\dag}\otimes U}_2\le\frac{1}{d}\norm{U^{\prime\dag}\otimes U^{\prime}-U^{\dag}\otimes U}_2.
\ee
Now evaluate $\norm{U^{\prime\dag}\otimes U^{\prime}-U^{\dag}\otimes U}_2$ recalling that $\tr(U^{\prime\dag}\otimes U^{\prime}U^{\prime}\otimes U^{\prime\dag})=d^2$:
\be
\frac{1}{d}\norm{U^{\prime\dag}\otimes U^{\prime}-U^{\dag}\otimes U}_2=\sqrt{2\left(1-\frac{1}{d^2}\tr(U^{\prime\dag}\otimes U^{\prime}U\otimes U^{\dag})\right)}\,,
\ee
\be
\frac{1}{d^2}\tr(U^{\prime\dag}\otimes U^{\prime}U\otimes U^{\dag})=\frac{1}{d^2}\tr(\hat{S}(U^{\prime\dag}\otimes U^{\prime})\hat{S}(U^{\dag}\otimes U))=\frac{1}{d^2}\frac{1}{\sqrt{\sum_{P,P^\prime\in\mathbb{Q}_{U}}\tr^2(U^\dag PUP^\prime)}}\sum_{P,P^\prime\in\mathbb{Q}_{U}}\tr^{2}(U^{\dag}PUP^{\prime})\,.
\ee
We are just left to the following series of inequalities:
\be
\sum_{P,P^\prime\in\mathbb{Q}_{U}}\tr^{2}(U^{\dag}PUP^{\prime})=d^4-\sum_{P,P^\prime\in\bar{\mathbb{Q}}_{U}}\tr^{2}(U^{\dag}PUP^{\prime})>d^4-\frac{\theta^2 d^4|\bar{\mathbb{Q}}_{U}|}{\card(U)}>d^4(1-\theta^2)\,,
\ee
where we used the fact that $\tr^{2}(U^{\dag}PUP^{\prime})< \theta \sqrt{\exp[M_{0}(\ket{U})]}$ iff $P\in\bar{\mathbb{Q}}_U$, where $\bar{\mathbb{Q}}_{U}$ is the complement set of $\mathbb{Q}_{U}$. Moreover, note that $|\bar{\mathbb{Q}}_U|=\card(U)-{\mathbb{Q}}_U<\card(U)$ where $\card(U):=|\{P,P^{\prime}\,|\,\tr(PUP^{\prime}U^{\dag})\neq0\}|$ and $M_{0}(\ket{U})=\log_2\frac{\card(U)}{d^2}$. We finally obtain that
\be
|\mathcal{F}_{U^\prime}-\mathcal{F}_{U}|\le \frac{1}{d}\norm{U^{\prime\dag}\otimes U^{\prime}-U^{\dag}\otimes U}_2\le \sqrt{2}\theta\,.
\ee
Now, $\mathcal{F}_{U^\prime}$ can be estimated in a similar fashion to $\mathcal{F}_{U}$ via Monte Carlo sampling, indeed,
\be
\mathcal{F}_{U^\prime}=\frac{1}{d^2}\sum_{\alpha}\tr(U^{\prime\dag}\otimes U^{\prime}A_{\alpha}\otimes A_{\alpha}^{\dag})=\frac{1}{d^4}\sum_{\mu,\nu}\tr(P_\mu\mathcal{U}(P_\nu))\tr(P_\mu U^{\prime}P_\nu U^{\prime\dag})=\aver{\mathcal{X}_{\mu \nu}^{\prime}}_{\Xi_{U^\prime}}
\ee
where $\mathcal{X}_{\mu\nu}^{\prime}:=\tr(P_\mu\mathcal{U}(P_\nu))/\tr(P_\mu U^{\prime}P_\nu U^{\prime\dag})$ and $\Xi_{U^\prime}$ is a probability distribution whose elements are:
\be
\Xi_{U^{\prime}}(P_\mu,P_\nu)=\begin{cases}\frac{\tr^2(P_\mu UP_\nu U^{\dag})}{\sum_{P,P^\prime\in\mathbb{Q}_{U}}\tr^2(PUP^\prime U^{\dag})}, \quad P_\mu,P_\nu\in\mathbb{Q}_U\\
0,\quad\text{otherwise}
\end{cases}
\ee
We can now define the estimator $\tilde{\mathcal{F}}_{U^\prime}$ of $\mathcal{F}_{U^\prime}$ in the usual way:
\be
\tilde{\mathcal{F}}_{U^{\prime}}=\frac{1}{k}\sum_{i}\tilde{\mathcal{X}_i}^{\prime}\,,
\ee
where $\tilde{X}_{i}^{\prime}=\frac{1}{\tr(P_{i}U^{\prime}P_{i}^{\prime}U^{\prime\dag})}\frac{1}{c_{i}(\mathcal{U})}\sum_{j=1}^{c_{i}(\mathcal{U})}\mathcal{P}_{ij}(\mathcal{U})$. We are ready to bound the probability to measure $\mathcal{F}_{U}$ with accuracy $\epsilon$ and find an upper bound to the resources $N_{\mathcal{U}}$:
\be
\operatorname{Pr}(|\mathcal{F}_{U}-\tilde{\mathcal{F}}_{U^\prime}|\le \epsilon)\ge1-\delta\,.
\ee
First $|\mathcal{F}_{U}-\tilde{\mathcal{F}}_{U^\prime}|\le |\mathcal{F}_{U}-{\mathcal{F}}_{U^\prime}|+|\mathcal{F}_{U^\prime}-\tilde{\mathcal{F}}_{U^\prime}|\le \sqrt{2}\theta+|\mathcal{F}_{U^\prime}-\tilde{\mathcal{F}}_{U^\prime}|$. Then, defining $m_{U}^\prime:=\min_{P}|\tr(P_iU^{\prime}P_jU^{\prime\dag})|/d$, since $\mathbb{E}\tilde{\mathcal{F}}_{U^\prime}={\mathcal{F}}_{U^\prime}$ we can use Hoeffding's inequality to bound the probability as
\be
\operatorname{Pr}(|\mathcal{F}_{U^\prime}-\tilde{\mathcal{F}}_{U^\prime}|\le\epsilon/2)\le 1-2\exp\left[-\frac{k\epsilon^2m_{U}^{\prime2}}{8}\right]\,.
\ee
Setting the probability to be greater than $1-\delta$, we find the necessary resources to be
\be
N_{\mathcal{U}}=\frac{8}{\epsilon^2m_{U}^{\prime2}}\ln(2/\delta)\,.
\ee
Setting $\theta\sqrt{2}=\epsilon/2$, we find $\operatorname{Pr}(|\mathcal{F}_{U}-\tilde{\mathcal{F}}_{U^\prime}|\le\epsilon)\ge1-\delta$. To complete the proof, it is necessary to lower bound $m_{U}^{\prime}$:
\be
m_{U}^{\prime}\ge \frac{\epsilon}{2\sqrt{2}}\sqrt{\exp[-M_{0}(\ket{U})]}\,,
\ee
which follows from Eq. \eqref{Ucutdef}. This concludes the proof.
\end{enumerate}

\subsection*{Proof of Corollary \hypers{dopedCliffordresources}{5}}\label{app:dopedCliffordresources}
In this section, we prove Corollary \hyperlink{dopedCliffordresources}{5}. Let us start with the lower bound of the number of resources needed for a doped Clifford circuit $C_{t}$---with associated Choi state $\ket{C_{t}}$---to be certified. From Theorem \hyperlink{theorem2}{4} we have:
    \be 
    N_{\mathcal{C}_{t}}\ge\frac{2}{\epsilon^{2}}\ln(2/\delta)\exp[M_{2}(\ket{C_{t}})]\,.
    \ee 
    To proceed we look at the average behavior of $\exp[M_{2}(\ket{C_{t}})]$:
    \be 
    \aver{(\exp[M_{2}(\ket{C_{t}})])}_{\mathcal{C}_{t}}=\aver{\left(\frac{d^6}{\sum_{P_1,P_2} \tr^{4}(P_1 U P_2 U^{\dag})}\right)}_{\mathcal{C}_{t}}\ge \frac{d^2}{\aver{\left[\tr(Q U^{\otimes 4} Q U^{\dag \otimes 4})\right]}_{\mathcal{C}_t}}\,,
    \ee 
    where $Q:=\frac{1}{d^2}\sum_{P\in\mathcal{P}_n}P^{\otimes 4}$, and we used the Jensen inequality to bound the average of $\aver{(\exp[M_{2}(\ket{C_{t}})])}_{\mathcal{C}_{t}}$. 
    To compute the average over doped Clifford circuits we use the techniques introduced in \cite{leone2021QuantumChaosQuantum,oliviero2021TransitionsEntanglementComplexity}, and obtain:
    \be 
    \aver{\left[\tr(Q U^{\otimes 4} Q U^{\dag \otimes 4})\right]}_{\mathcal{C}_{t}}\!\!=\!\left[\frac{4(6-d^2+d^4)}{d^2(d^2-9)}+(d^2-1)\left(\frac{(d+2)(d+4)f_{+}^t}{6 d (d+3)}+\frac{(d-2)(d-4)f_{-}^{t}}{6 d (d-3)}+\frac{(d^2-4)\left(\frac{(f_{+}+f_{-})}{2}\right)^{t}}{3 d^2}\right)\right]^{-1}\,,
    \ee 
    where $f_{\pm}=\frac{3d^2\mp 3d -4}{4(d^2-1)}$. One easily shows that $\aver{\left[\tr(Q U^{\otimes 4} Q U^{\dag \otimes 4})\right]}_{\mathcal{C}_{t}}=\Theta(\exp[t\log_24/3])$, and thus the number of resources is lower bounded by:
    \be 
    N_{\mathcal{C}_{t}}\ge\Theta(\exp[t\log_24/3])\,.
    \ee 
    To prove the upper bound to the number of resources we use the upper bound in Theorem \hyperlink{theorem2}{4}. 
    \be \label{eqbound}
    N_{\mathcal{C}_{t}}\le \frac{64}{\epsilon^{4}}\ln(2/\delta)\exp[\nu(U)]\,.
    \ee 
    As proven in \cite{jiang2021LowerBoundTcount} the unitary stabilizer nullity can be upper bounded with the \textit{T-count} $t(U)$, which corresponds to the minimum number of T gates required to implement the unitary $U$. Equation \eqref{eqbound} can be upper bounded via the \textit{T count} as
    \be 
     N_{\mathcal{C}_{t}}\le \frac{64}{\epsilon^{4}}\ln(2/\delta)\exp[\nu(U)]\le \frac{64}{\epsilon^{4}}\ln(2/\delta)\exp[t]\simeq \Theta(\exp[t])\,,
    \ee 
    where we used that for doped Clifford circuits $t(U)=t$. This concludes the proof. 

    \section{Shadow fidelity estimation}\label{app:shadowfidestimation}
Let $N_{\tilde{\rho}}$ be the number of physical preparation of $\rho$. Let $N_{\tilde{\psi}}=k\times l$, where $k$ is the number of Clifford circuits drawn uniformly at random from the Clifford group and $l$ is the number of realizations of the experiment. For a single experiment, we have
\be
\mathcal{F}_{s}=\frac{1}{k}\sum_{C_i\in\mathbb{C}_s}[(d+1)\braket{\bar{x}_i|C_{i}^{\dag}\rho C_i|\bar{x}_i}-1]\,.
\ee
Defining $\overline{\mathcal{F}}=\operatorname{median}\{\mathcal{F}_s~|~s=1,\ldots, l\}$, we already know that~\cite{kliesch2021TheoryQuantumSystem}
\be
\operatorname{Pr}[|\mathcal{F}(\rho,\tilde{\rho})-\overline{\mathcal{F}}|\le \epsilon]\ge 1-\delta
\ee
for $N_{\tilde{\rho}}\ge\frac{160}{\epsilon^2}\ln \delta^{-1}$ and $k=8\ln 8\delta^{-1}$. Now, let $\rho_{\bar{x}_i}$ be the classical estimation of the outcome probability $\braket{\bar{x}_i|C_{i}^{\dag}\rho C_i|\bar{x}_i}$. Let $N_{cl}$ be the classical resources necessary for ensuring
\be
\operatorname{Pr}[|\rho_{\bar{x}_i}-\braket{\bar{x}_i|C_{i}^{\dag}\rho C_i|\bar{x}_i}|\le \epsilon]\ge 1-\delta\,.
\ee
Then, defining 
\be
\tilde{\mathcal{F}}_{s}=\frac{1}{k}\sum_{C_i\in\mathbb{C}_s}[(d+1)\rho_{\bar{x}_i}-1]\,
\ee
one has
\be
\operatorname{Pr}[|\tilde{\mathcal{F}}_{s}-\mathcal{F}_{s}|\le (d+1)\epsilon]\ge 1-\delta\,.
\ee

\end{document}